\documentclass[a4paper, notoc]{JHEP3}
\usepackage[dvips]{graphicx}
\usepackage{amsfonts}
\usepackage{amssymb}
\usepackage{epsfig}
\usepackage{cite}

\newcommand{\be}{\begin{equation}}
\newcommand{\ee}{\end{equation}}
\newcommand{\bea}{\begin{eqnarray}}
\newcommand{\eea}{\end{eqnarray}}
\newcommand{\mbb}{\mathbb}
\newcommand{\ti}{\times}
\newcommand{\half}{\frac{1}{2}}
\newcommand{\mc}{\mathcal}

\newcommand{\beqa}{\begin{eqnarray}}
\newcommand{\eeqa}{\end{eqnarray}}

 \newlength{\wth}
 \newcommand{\twographs}[2]{%
 \unitlength=1.1in
 \begin{picture}(5.8,2.6)
 \put(0,0){\epsfig{file=#1.eps, width=0.8\wth}}
 \put(2.5,0){\epsfig{file=#2.eps, width=0.8\wth}} 
 \put(0,2.1){(a)}
 \put(2.3,2.1){(b)}
 \end{picture}
}

 \newcommand{\twographsbig}[2]{%
 \unitlength=1.1in
 \begin{picture}(5.8,2.6)
 \put(0,0){\epsfig{file=#1.eps, width=0.95\wth}}
 \put(2.5,0){\epsfig{file=#2.eps, width=0.95\wth}}
 \put(0,2.1){(a)}
 \put(2.3,2.1){(b)}
 \end{picture}
}

 \newcommand{\fourgraphs}[4]{%
 \unitlength=1.1in
 \begin{picture}(6.,5)
 \put(0,0){\epsfig{file=#3, width=\wth}}
 \put(2.5,0){\epsfig{file=#4, width=\wth}}
 \put(0,2.6){\epsfig{file=#1, width=\wth}}
 \put(2.5,2.6){\epsfig{file=#2, width=\wth}}
 \put(0,2.1){(c)}
 \put(3.0,2.1){(d)}
 \put(0,4.7){(a)}
 \put(3.0,4.7){(b)}
 \end{picture}
}

 \title{Soft SUSY Breaking Terms for Chiral Matter in IIB String Compactifications}
\author{Joseph P. Conlon, Shehu S. Abdussalam, Fernando Quevedo, Kerim Suruliz \\ DAMTP,
  Centre for Mathematical
Sciences,\\
  Wilberforce Road, Cambridge, CB3 0WA, UK.\\
 E-mail:
  \email{s.s.abdussalam@damtp.cam.ac.uk} , \email{j.p.conlon@damtp.cam.ac.uk} ,
  \email{f.quevedo@damtp.cam.ac.uk}, \email{k.suruliz@damtp.cam.c.uk}}

\abstract{This paper develops the computation of soft supersymmetry
breaking terms for chiral D7 matter fields in IIB Calabi-Yau flux
compactifications with stabilised moduli. We determine explicit
expressions for soft terms for the single-modulus KKLT scenario and
the multiple-moduli large volume scenario. In particular we use the
chiral matter metrics for Calabi-Yau backgrounds recently computed
in hep-th/0609180. These differ from the better understood metrics
for non-chiral matter and therefore give a different structure of
soft terms. The soft terms take a simple form depending
explicitly on the modular weights of the corresponding matter
fields. For the large-volume case we find that in the simplest D7
brane configuration, scalar masses, gaugino masses and A-terms are
very similar to the dilaton-dominated scenario. Although all soft
masses are suppressed by $\ln(M_P/m_{3/2})$ compared to the
gravitino mass, the anomaly-mediated contributions {\it do not}
compete, being doubly suppressed and thus subdominant to the 
gravity-mediated tree-level terms. Soft terms are flavour-universal to leading
order in an expansion in inverse K\"ahler moduli. They also do not
introduce extra CP violating phases to the effective action.  We argue
that soft 
term flavour universality should be a property of the large-volume
compactifications, and more generally IIB flux models, in which
flavour is determined by the complex structure moduli while
supersymmetry is broken by the K\"ahler moduli. For the simplest
large-volume
case we run the soft terms to low energies and present some sample
spectra and a basic phenomenological analysis.}

\preprint{DAMTP-2006-89 \\ hep-th/0610129}

\begin{document}

\tableofcontents

\section{Introduction}

The imminent advent of the LHC is an excellent motivation to develop techniques to
relate high energy string compactifications to observable low-energy collider physics.
The LHC will be an unprecedented probe of the terascale and of the physics that stabilises the
electroweak hierarchy.
If supersymmetry is discovered at the LHC and the spectrum of
superpartners measured, it will be necessary to
connect the features of this spectrum to a more fundamental
theory such as string theory. This connection is provided by the
soft terms, that parametrise supersymmetry breaking and whose RG running determines the low-scale spectrum.

Supersymmetry breaking and the effective supergravity Lagrangian are among the most studied areas of string
phenomenology \cite{softreview, bim}. 
There are two principal tasks here, the first to determine how supersymmetry is broken and
the second to determine how this breaking is transmitted to the
matter fields. The first task primarily requires control of the moduli potential.
This determines the vacuum structure of the theory. By studying the minima of the potential,
it is possible to work out which fields break supersymmetry and the magnitude of the relevant
F-terms. The second task requires a knowledge of the couplings of the fields breaking supersymmetry to
the matter sector. In a supergravity context, this is equivalent to knowing the modular dependence of
the gauge kinetic functions and matter metrics. These are necessary to go from the moduli F-terms to
soft masses for squarks and sleptons.

In recent years much progress has been made on the first problem. In the original studies of soft terms in
string compactifications, it was necessary to classify supersymmetry
breaking as $S$-, $T$- or $U$-dominated, depending on the dominant F-term,
without attempting to specify the underlying microscopic source of supersymmetry
breaking \cite{bim}. These studies were restricted to the heterotic string and were limited
by the lack of controlled potentials in which all moduli were stabilised.
In recent years much progress has been made in the field of
moduli stabilisation \cite{hepth0105097, hepth0301240, hepth0509003},
particularly in the context of type II compactifications.
The most substantial advance is that it is now possible to compute
potentials that can stabilise all the moduli in a controlled fashion.
By studying the vacuum structure of these
potentials, it becomes possible to determine from first
principles how supersymmetry is broken and by which moduli.

Progress has also been made on the second problem. At a practical level, this requires knowledge
of the gauge kinetic functions $f_a(\Phi_i)$ and the matter metrics $\tilde{K}_{\alpha \bar{\beta}}(\Phi_i)$.
The former, being holomorphic, are relatively easy to compute but it is in general a hard problem to compute the
latter.
However both are necessary to go from a supersymmetry-breaking vacuum to soft masses and trilinear A-terms.
On flat toroidal backgrounds, the K\"ahler metrics for chiral bifundamental matter
can be computed through explicit string scattering computations
\cite{hepth0404134, lerda}. In Calabi-Yau backgrounds, it is possible to use dimensional reduction to compute K\"ahler metrics for
D7 adjoint and Wilson line matter, but this does not apply to
bifundamental matter. However it was recently described,
in a companion paper \cite{CCQPaper} to the present one, how to
compute bifundamental K\"ahler metrics for
Calabi-Yau backgrounds. The purpose of this paper is to apply these results 
to the computation of soft terms.\footnote{The soft terms for chiral
matter fields in toroidal models were considered in \cite{hepth0406092, ibanezfont}.}.

For realistic models of supersymmetry breaking, it is necessary that
the scale of supersymmetry breaking be hierarchically small, in
order to generate a TeV-scale gravitino mass and explain the
hierarchy between the weak and Planck scales. 
One class of models in
which this occurs naturally are the large-volume models of
\cite{hepth0502058, hepth0505076}. 
These arise in IIB flux compactifications once the leading
$\alpha'$ correction is used to lift the no-scale structure. 
They fix all the moduli and
are characterised by the resulting exponentially large volume
$\mc V $. The large volume arises from the combination of $\alpha'$
K\"ahler corrections and non-perturbative superpotential corrections.
The
large volume lowers both the string scale $m_s$ and the gravitino
mass $m_{3/2}$,
$$
m_{s} \sim \frac{M_P}{\sqrt{\mc V}}\, , \qquad m_{3/2} \sim
\frac{M_P}{\mc V} W_0.
$$
Here $M_P$ is the Planck mass and $W_0$ the flux-induced
superpotential. For generic values of $W_0 \sim 1$, a volume
$\mc{V} \sim 10^{15} l_s^6$
corresponds to TeV-scale soft terms. This gives a string scale $m_s
\sim 10^{11} \hbox{GeV}$. A notable feature of these models
is that the large volume occurs dynamically and there is no need to
fine-tune $W_0$.
If we do fine-tune the flux
superpotential $W_0 \ll 1$, the value of the string
scale can be increased to the GUT scale. 
Fine-tuned flux superpotentials are characterstic of the
KKLT scenario, for which
$W_0\sim 10^{-13}$ gives rise to a TeV gravitino mass and a GUT
string scale.

Supersymmetry breaking and soft terms for the large volume models have been
discussed in \cite{hepth0505076, aqs, hepth0605141}. The first of these
papers computed the overall scale of the soft terms and the second
identified a logarithmic suppression in the gaugino masses compared
to the gravitino mass, with the scalar masses generically not
suppressed. However these results were for non-chiral D7 matter
fields which are less relevant for phenomenology. Chiral matter was
not considered in \cite{hepth0605141} 
due to the fact that the appropriate K\"ahler potentials were not
known.

An alternative scenario of soft terms has also been put forward in
\cite{choi, falkowski,mirage}. This
uses the simplest KKLT scenario with one K\"ahler
modulus.\footnote{Actually the separation of these as two alternative
scenarios is not really appropriate. The reason is that
starting with the minimum of the  scalar potential in the large
volume scenario, which has a natural $\mc{O}(1)$ flux superpotential
$W_0$, 
it has been shown that, reducing the value of $W_0$, in
the limit of $W_0 \ll 1$ one reproduces the KKLT minimum that requires
very small $W_0.$ It is instead more appropriate to talk about two limits of
the same scenario.} A suppression of all soft terms was also found
in such a way that the anomaly-mediated contribution to the soft terms
was of a similar order to the soft terms induced by the moduli
F-terms. This gave rise to an 
interesting scenario that has been named
mirage mediation. Having a single modulus simplifies some of the
calculations and the structure of soft terms was found for all
possible modular weights of the D7 matter fields. However the
phenomenological analysis was performed only for the modular weights
corresponding to non-chiral matter, again because those for chiral
fields were unknown.

 In \cite{CCQPaper} the functional dependence 
of the chiral matter metric on the moduli fields was recently computed
for the above models, and
in this paper we will use these results to further develop the
computation of soft terms. We will then obtain the expressions for
soft terms for chiral fields in both scenarios. These are the
phenomenologically relevant expressions since squarks and sleptons
come from chiral matter fields.

Both the above models involve gravity-mediated supersymmetry breaking.
A common criticism of gravity-mediation
 is that generic supersymmetry
breaking leads to non-universal squark masses and large unobserved
flavour-changing neutral currents. This arises because in gravity
mediation both flavour and supersymmetry breaking
are UV-sensitive, and thus one would expect that the two cannot be decoupled -
the physics breaking supersymmetry is also sensitive to
flavour. We will consider this point and argue that it does not hold in the large-volume models,
or more generally in IIB flux compactifications, as
the sector that gives flavour - the complex structure moduli - and
the sector that breaks supersymmetry - the K\"ahler moduli - are
decoupled at leading order.

The structure of this paper is as follows. In section \ref{sec2} we review
the basics of gravity-mediated supersymmetry breaking and recall the
results of \cite{CCQPaper} on bifundamental matter metrics.
In section \ref{largevol}
we start by reviewing the large volume model and its geometry. In section
\ref{secstc} we use these matter metrics to
perform the computations of soft terms for diagonal matter metrics, 
in the limit of large cycle volume
and dilute fluxes. In section \ref{Nondiagonal} we repeat these computations for 
the case of non-diagonal matter metrics, obtaining identical results. In section \ref{Universality}
we give a general discussion of
universality in gravity-mediated models and explain why the decoupling of K\"ahler and complex
structure moduli naturally makes the soft terms of IIB flux models flavour-universal.
We also compute the
anomaly mediated contribution within supergravity and find it to be
suppressed compared to the tree-level soft terms.
In section \ref{RGrunning} we perform a
simple phenomenological analysis of the soft terms, carrying out the
running and illustrating this with sample spectra. In section
\ref{kkltl} we discuss the simplest KKLT one-modulus scenario, mostly
  to illustrate how our results modify the previous analysis of this
  scenario. Finally, in section
\ref{Conclusions} we conclude.

\section{Moduli SUSY Breaking}
\label{sec2}

\subsection{Effective $N=1$ Supergravity Lagrangian}

A four dimensional $N=1$ supergravity Lagrangian is specified at two
derivatives by the K\"ahler potential $K$, superpotential $W$ and gauge
kinetic function $f_a$.
The computation of soft terms starts by expanding these as a power series
in the matter fields, \bea \label{PowerSeriesExpansion} W & = &
\hat{W}(\Phi) + \mu(\Phi) H_1 H_2 + \frac{1}{6} Y_{\alpha
  \beta \gamma}(\Phi) C^{\alpha} C^{\beta} C^{\gamma} + \ldots, \\
\label{MatterK} K & = & \hat{K}(\Phi, \bar{\Phi}) +
\tilde{K}_{\alpha \bar{\beta}} (\Phi, \bar{\Phi}) C^{\alpha}
C^{\bar{\beta}} + \left[ Z(\Phi,
  \bar{\Phi}) H_1 H_2 + h.c. \right] + \ldots, \\
f_a & = & f_a(\Phi). \eea $C^\alpha$ denotes a matter field and we
have here, for convenience, separated the Higgs fields $H_{1,2}$
from the rest of the matter fields
 and specialised to the
MSSM by assuming two Higgs doublets. We use $\Phi$ to denote an
arbitrary modulus field and do not specify the total number of
moduli.

In IIB compactifications
the K\"ahler potential and superpotential for the moduli take the
standard form \cite{hepth0301240, hepth9906070, hepth0204254,
hepth0403067}, \bea \hat{K}(\Phi, \bar{\Phi}) & = & - 2 \ln \left(
\mc{V} + \frac{\hat{\xi}}{2 g_s^{3/2}} \right) - \ln \left(
i \int \Omega \wedge \bar{\Omega} \right) - \ln (S + \bar{S}). \\
\hat{W}(\Phi) & = & \int G_3 \wedge \Omega + \sum_i A_i e^{-a_i
T_i}. \eea $\mc{V}$ is the Einstein-frame volume of the Calabi-Yau.
The first term in $\hat{K}$ is the K\"ahler moduli dependence,
including the leading $\alpha'$ correction, while the second and third give
the complex structure and dilaton dependence. In the
superpotential, the first term is the flux-induced superpotential
\cite{hepth9906070} that
depends on dilaton and complex structure moduli. 
When evaluated at the minimum of the potential with respect
to these moduli, this is denoted by $W_0$. The second term is the nonperturbative
superpotential responsible for fixing the K\"ahler moduli. Note
the separate dependences of $\hat{W}$ and $\hat{K}$ on the
complex and K\"ahler moduli. This will play an important role when we
subsequently discuss flavour universality.

The gauge kinetic functions $f_a(\Phi)$ depend on whether the
gauge fields come from D3 or D7 branes and, in the latter case, on the
4-cycle wrapped by the D7 brane. If $T_i$ is the K\"ahler modulus
corresponding to a particular 4-cycle, reduction of the DBI action
for an unmagnetised brane wrapped on that cycle gives \be
\label{GaugeCoupling} f_i = \frac{T_i}{2 \pi}. \ee We are interested
in magnetised branes wrapped on 4-cycles. 
The magnetic fluxes alter
(\ref{GaugeCoupling}) to \be \label{FluxGaugeCouplings} f_i =
h_i(F)\, S + \frac{T_i}{2 \pi}, \ee where $h_i$ depends on the fluxes
present on that stack. This can be understood microscopically. The
gauge coupling is given by an integral over the cycle $\Sigma$
wrapped by the brane, \be \frac{1}{g^2} = \int_{\Sigma} d^4 y \,
e^{-\phi} \sqrt{g + (2 \pi
  \alpha')\mc{F}}.
\ee Thus in the presence of flux, the gauge coupling is no longer
determined simply by the cycle volume. The factors $h_a(F)$ have
been explicitly computed for toroidal orientifolds
\cite{hepth0404134, hepth0406092}.

The presence of the term $h_a(F)$ can also be understood from the
mirror IIA description. In the mirror description, IIB D7 branes
become IIA D6 branes (see for example \cite{hepth0502005}). The different 
choices of magnetic flux on D7-branes correspond to different cycles 
wrapped by D6 branes. As different cycles have different sizes, it is
 manifest that the
different branes should have different gauge couplings. The
functions $h_a(F)$ are topological and so in principle may be
computed even on Calabi-Yau geometries. However a sensible
computation would require a geometry already yielding the Standard Model
matter content - a hard and unresolved problem.
Notice that (\ref{FluxGaugeCouplings}) allows for non-unified gauge
couplings, which are consistent
 with an intermediate string scale.

 The computation of soft scalar masses and
trilinears depends crucially on the K\"ahler potential for matter
fields $\tilde{K}_{\bar{\alpha} \beta} (\Phi, \bar{\Phi}) $ and
$Z(\Phi,
  \bar{\Phi})$. However these are also the
least understood part of the supergravity effective action, as being
nonholomorphic they are not protected in the same way that the
superpotential is.
For D3 and D7 fields living in the adjoint of the gauge group the
matter metrics have been computed in \cite{louis}, and for chiral
bifundamental matter fields the K\"ahler potential has been fully
computed only in particular toroidal orientifold models \cite{hepth0406092}.

This paper is a companion to \cite{CCQPaper}, where it was recently
described how for the geometry of the large-volume models the
dependence of the bifundamental matter metric on the K\"ahler moduli
may be computed using scaling arguments involving
 the physical Yukawa couplings,\footnote{Formula (\ref{PhysicalYukawaCouplings})
specialises to the case of diagonal matter metrics,
$\tilde{K}_{\alpha\beta} = \tilde{K}_\alpha \delta_{\alpha\beta}$ (no
sum on right hand side), but the results hold equally well in the
general, non-diagonal case.}
\be \label{PhysicalYukawaCouplings} \hat{Y}_{\alpha \beta \gamma}  =
e^{\hat{K}/2} \frac{Y_{\alpha \beta \gamma}}{ (\tilde{K}_\alpha
\tilde{K}_\beta \tilde{K}_\gamma)^{\half}}. \ee As the combination
of shift symmetry and holomorphy implies that the K\"ahler moduli do not
appear perturbatively in $Y_{\alpha \beta \gamma}$, it is possible to extract the
behaviour of the matter metrics $\tilde{K}_{\alpha
\bar{\beta}}(T_i)$ through the scaling of the physical
  Yukawa couplings $\hat{Y}_{\alpha \beta \gamma}(T_i)$. This scaling
  may be computed through wavefunction overlap or direct
  dimensional reduction of higher-dimensional Yang-Mills.

Let us review the simplest argument of \cite{CCQPaper}. 
The notation used here will be more fully explained in section \ref{largevol}.
We suppose
the Standard Model is supported on a small cycle $\tau_s$ within
a large bulk $\tau_b$. This geometry is that encountered in models of
branes at (resolved) singularities or the large volume models which
will be reviewed in section \ref{largevol}. In this case locality
implies the physical Yukawa couplings are determined only by the local
geometry and are independent of the overall volume. As $\hat{K} = -2
\ln {\cal{V}}$, for this to apply in (\ref{PhysicalYukawaCouplings}) we must
have
\be
\tilde{K}_{\alpha} \sim \frac{k_{\alpha \bar{\beta}}(\phi, \tau_s)}{\mc{V}^{2/3}},
\ee
where $\phi$ are complex structure moduli.
Expanding $k_{\alpha \bar{\beta}}(\phi, \tau_s)$ in a power series in $\tau_s$,
we can write the resulting metric as 
\be 
\label{MatterMetric} 
\tilde{K}_{\alpha \bar{\beta}} =
\frac{\tau_s^{\lambda}}{\mc{V}^{2/3}}k_{\alpha \bar{\beta}}(\phi),
\ee 
This expression holds
in the limit of dilute fluxes and large cycle volume $\tau_s$ and
will receive corrections subleading in $\tau_s$. For the minimal
model in which all branes wrap the same cycle, it was shown in
\cite{CCQPaper} that $\lambda = 1/3$. For
other cases $\lambda$ may take values between 0 and 1.
However in computations we will often start by using the more general form
\be
\label{FactorisedK} \tilde{K}_{\alpha \bar{\beta}} = \frac{k_{\alpha \bar{\beta}}(\tau_s,
  \phi)}{\tau_b^{p}}.
\ee 
The purpose of this is to illustrate the special cancellations that
occur uniquely for the form (\ref{MatterMetric}).

For KKLT models with a single modulus, similar arguments
\cite{CCQPaper} allow us to likewise write
\be
\label{KKLTK}
\tilde{K}_{\alpha \bar{\beta}} = \frac{k_{\alpha \bar{\beta}}(\phi)}{\tau^{2/3}},
\ee
where $\tau = \hbox{Re}(T)$ is the size of the single 4-cycle.

We refer
to \cite{CCQPaper} for the full derivations of (\ref{MatterMetric}) and
(\ref{KKLTK}), as our focus is 
on using these formulae to compute soft terms.

\subsection{Soft Breaking Terms}
\label{subsec22}

 The full $\mc{N}=1$ scalar potential is \be
\label{n=1pot} V = e^K \left( K^{i \bar{j}} D_i W D_{\bar{j}}
\bar{W} - 3 \vert W \vert^2 \right), \ee 
where $D_iW= \partial_i W +
(\partial_i K)W$. 
The soft supersymmetry breaking terms are found by
expanding
 (\ref{n=1pot}) in powers of the matter fields using the expansions (\ref{PowerSeriesExpansion}).
We will give the formulae for these below for
diagonal matter metrics and in section \ref{Nondiagonal} for nondiagonal matter
metrics, but first we make some general remarks.

Gravity-mediated supersymmetry breaking is quantified through the
moduli F-terms, given by \be F^m = e^{\hat{K}/2} \hat{K}^{m \bar{n}}
D_{\bar{n}} \bar{\hat{W}}. \ee We can assume without loss of
generality that the superpotential $\hat{W}$ is real. To see this,
note from the Lagrangian in Appendix G of \cite{WessBagger} that a
phase rotation of the superpotential can be achieved through phase
redefinitions of the gauginos, chiral fermions and gravitino. In
this case the gravitino mass is real while both the gaugino masses
and A-terms are generically complex.

Given the F-terms, the easiest soft parameters to compute are the
gaugino masses. In terms of the unnormalised field $\lambda^a$, the
canonically normalised gaugino field $\hat{\lambda}^a$ is \be
\hat{\lambda}^a = (\hbox{Re} f_a)^\half \lambda^a. \ee The
canonically normalised gaugino masses are then given by \be
\label{CNGM} M_a = \half \frac{F^m \partial_m f_a}{ \hbox{Re} f_a}.
\ee The gaugino masses that follow from (\ref{FluxGaugeCouplings})
are given by \be M_i = \frac{F^s}{2} \frac{1}{T_s + 2 \pi h_a(F) S}.
\ee In the limit of large cycle volume, the flux becomes
diluted and the gauge coupling is determined solely by the cycle
size. We shall mostly work in this dilute flux approximation, in
which the gaugino masses become \be \label{GauginoMass} M_i =
\frac{F^s}{2 \tau_s}. \ee The fractional non-universality of gaugino
masses is set by the flux contribution to the gauge couplings. The
quasi-universal relation (\ref{GauginoMass})
 holds for the minimal geometry: if there are several small cycles
 involved the expressions for gaugino masses may involve several
moduli and be more complicated.

For a diagonal matter metric the soft scalar Lagrangian can be
written as \be \mc{L}_{soft} = \tilde{K}_{\alpha} \partial_\mu
C^{\alpha}
\partial^\mu \bar{C}^{\bar{\alpha}} -
m_\alpha^2 C^{\alpha} \bar{C}^{\bar{\alpha}} - \left(
\frac{1}{6} A_{\alpha \beta \gamma} \hat{Y}_{\alpha \beta \gamma}
C^\alpha C^\beta C^\gamma + B \hat{\mu} \hat{H}_1
\hat{H}_2 + h.c. \right), \ee with the scalar masses, A-terms and
B-term being given by \cite{bim}
\bea \label{SoftMassDiagFormula} m_\alpha^2  &
=  & (m_{3/2}^2 + V_0) - F^{\bar{m}}F^n \partial_{\bar{m}}
\partial_n \log \tilde{K}_\alpha. \\
\label{ATermFormula} A_{\alpha \beta \gamma} & = & F^m \left[
\hat{K}_m + \partial_m \log
  Y_{\alpha \beta \gamma} - \partial_m \log (\tilde{K}_\alpha \tilde{K}_\beta \tilde{K}_\gamma) \right]. \\
\label{BTermFormula} B \hat{\mu} & = & (\tilde{K}_{H_1}
\tilde{K}_{H_2})^{-\half} \Bigg\{  e^{\hat{K}/2} \mu \left( F^m
\left[\hat{K}_m + \partial_m \log \mu - \partial_m \log
  (\tilde{K}_{H1} \tilde{K}_{H2}) \right] - m_{3/2} \right) \nonumber \\
& &  + \left( 2 m_{3/2}^2 + V_0 \right) Z - m_{3/2}
\bar{F}^{\bar{m}} \partial_{\bar{m}} Z + m_{3/2} F^m  \left[
\partial_m Z -
Z \partial_m  \log (\tilde{K}_{H_1} \tilde{K}_{H_2})\right] - \nonumber \\
& & \bar{F}^{\bar{m}} F^n \left[ \partial_{m} \partial_n Z -
(\partial_{\bar{m}} Z) \partial_n \log (\tilde{K}_{H_1}
\tilde{K}_{H_2})\right] \Bigg\}. \eea 

There exist similar formulae for non-diagonal matter metrics that we
will present in section \ref{Nondiagonal}.
The computation of soft scalar masses requires knowledge of the
matter metric, which we take from (\ref{MatterMetric}) and
(\ref{FactorisedK}). 
As above these
expressions should be understood as holding in the limit of large
cycle volumes and dilute fluxes. We will perform our computation of
MSSM soft parameters for both diagonal and non-diagonal matter
metrics, before subsequently explaining why the latter reduces to
the first and how the decoupling of K\"ahler and complex structure
moduli gives rise to soft term universality.

In order to compute the soft
breaking terms - gaugino and scalar masses, A-terms and the B-term,
we need a class of models for which we can compute the corresponding
F-terms. We can then combine these and the expression for
$\tilde{K}_{\alpha \bar{\beta}}$ into the above formulae to compute
these quantities.
We will first discuss the
structure of soft terms in the large volume scenario which we need
to introduce in detail. We later discuss the implications of our
matter metric for the one-modulus KKLT scenario.

\section{The Large-Volume Model}
\label{largevol}

\subsection{Geometry}

We start this section with a brief description of the geometry of the
large-volume models.
These models exist within the framework of IIB flux compactifications
with D3 and D7 branes.

 The dilaton and complex structure moduli are stabilised by fluxes. The K\"ahler moduli are
stabilised by a combination of $\alpha'$ corrections and nonperturbative superpotentials.
As shown in \cite{hepth0502058, hepth0505076}, these very generally interact
to produce one exponentially large cycle controlling the
overall volume together with $h^{1,1} - 1$ small cycles.
We denote
the
large and small moduli by
$T_b = \tau_b + i c_b$ and $T_i = \tau_i + i c_i$ respectively, with $i = 1 \ldots
h^{1,1} -1$. Consistent with this, we assume the volume can be written as
\be
\mc{V} = \tau_b^{3/2} - h(\tau_i),
\ee
where $h$ is a homogeneous function of the $\tau_{i}$ of degree 3/2.

The large volume lowers both the string
 scale and gravitino mass,
$$
m_s \sim \frac{M_P}{\sqrt{\mc{V}}} \qquad \hbox{ and } \qquad m_{3/2} \sim
m_{soft} \sim \frac{M_P}{\mc{V}}.
$$
The stabilised volume is exponentially sensitive to the stabilised
string coupling, $\mc{V} \sim e^{\frac{c}{g_s}}$, and
may thus take arbitrary values.
A volume $\mc{V} \sim 10^{15} l_s^6 \equiv 10^{15} (2\pi \sqrt{\alpha'})^6$ is required to explain the weak/Planck
hierarchy and give a TeV-scale gravitino mass. As $m_s \gg m_{3/2}$,
the low-energy phenomenology is that of the MSSM and thus the
computation of soft terms is central to the study of the low energy phenomenology.

The simplest model is that of $\mbb{P}^4_{[1,1,1,6,9]}$, which we use
as our working example. For this the volume can be written as \cite{DenefDouglasFlorea}
\be
\mc{V} = \frac{1}{9\sqrt{2}} \left( \tau_b^{3/2} - \tau_s^{3/2}
\right).
\ee
$\tau_b$ and $\tau_s$ denote the big and small cycles. Moduli
stabilisation for this geometry has been studied very explicitly in
\cite{hepth0502058, hepth0505076, hepth0605141}, and we shall not
repeat this discussion here.
Our prime interest here is the computation of soft terms and the
analysis of universality.
The geometry of the large volume models is illustrated in
figure \ref{LargeVolumePic}.
\FIGURE{\makebox[15cm]{\epsfxsize=15cm \epsfysize=12cm
\epsfbox{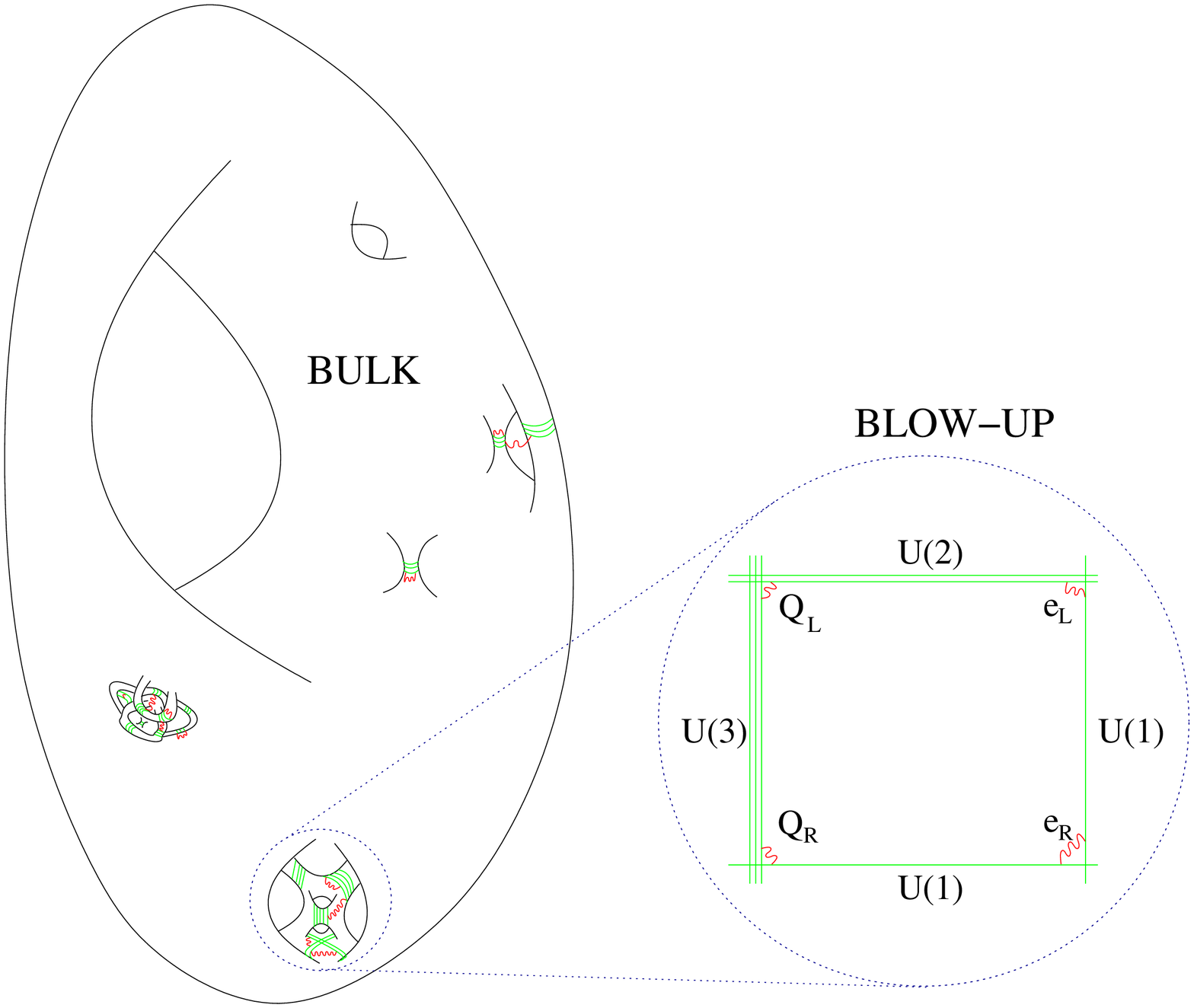}}\caption{The physical picture: Standard Model matter is supported on a
small blow-up cycle located within the bulk of a very large
Calabi-Yau. The volume of the Calabi-Yau sets the gravitino mass and 
is responsible for the weak/Planck hierarchy.}\label{LargeVolumePic}}
The stabilised volumes of the small cycles are $\tau_s \sim \ln
\mc{V}$. For $\mc{V} \sim 10^{15}$,\footnote{If no dimensions are
  given the volume can be assumed to be measured in string units.} D7-branes
wrapped on such cycles have gauge
couplings qualitatively similar to those of the Standard Model.
If the branes are magnetised, Standard
Model chiral matter can arise from strings stretching between stacks
of D7 branes. We assume the Standard Model arises from a stack of 
magnetised branes wrapping one (or more) of the small cycles. We do
not attempt to specify the local geometry - for work in this
direction, see \cite{BranesAtSingularities}.

In what we call the `minimal model', there exists only one small
blow-up 4-cycle on which a stack of magnetised
D7 branes are wrapped. For the $\mbb{P}^4_{[1,1,1,6,9]}$ model, this
corresponds to wrapping D7 branes on the small cycle.
The existence of only one small cycle need not be incompatible with the several
different gauge factors and couplings of the Standard Model. The spectrum of chiral
fermions depends on the magnetic flux $F$ present on the brane
worldvolume. This is quantised on 2-cycles $\Sigma_i$,
\be
\int_{\Sigma_i} F \in \mbb{Z}.
\ee
If several such
2-cycles exist within the 4-cycle, different brane stacks can be realised through
different choices of 2-form flux on these 2-cycles. This is consistent
with there being only one small Calabi-Yau 4-cycle,
as these 2-cycles may be homologically trivial within the Calabi-Yau and only non-trivial when
restricted to the 4-cycle.

There is also no obstruction to the presence of chiral matter -
if the cycle is a blow-up cycle, it is by construction localised in the Calabi-Yau.
The branes thus cannot move off the cycle and there is no adjoint
matter to give masses to strings stretching between branes.
This permits a chiral spectrum, as is found in explicit models of branes
at singularities.
The geometry of this minimal model is shown in figure
\ref{MinimalModel}.
\FIGURE{\makebox[15cm]{\epsfxsize=15cm \epsfysize=15cm
\epsfbox{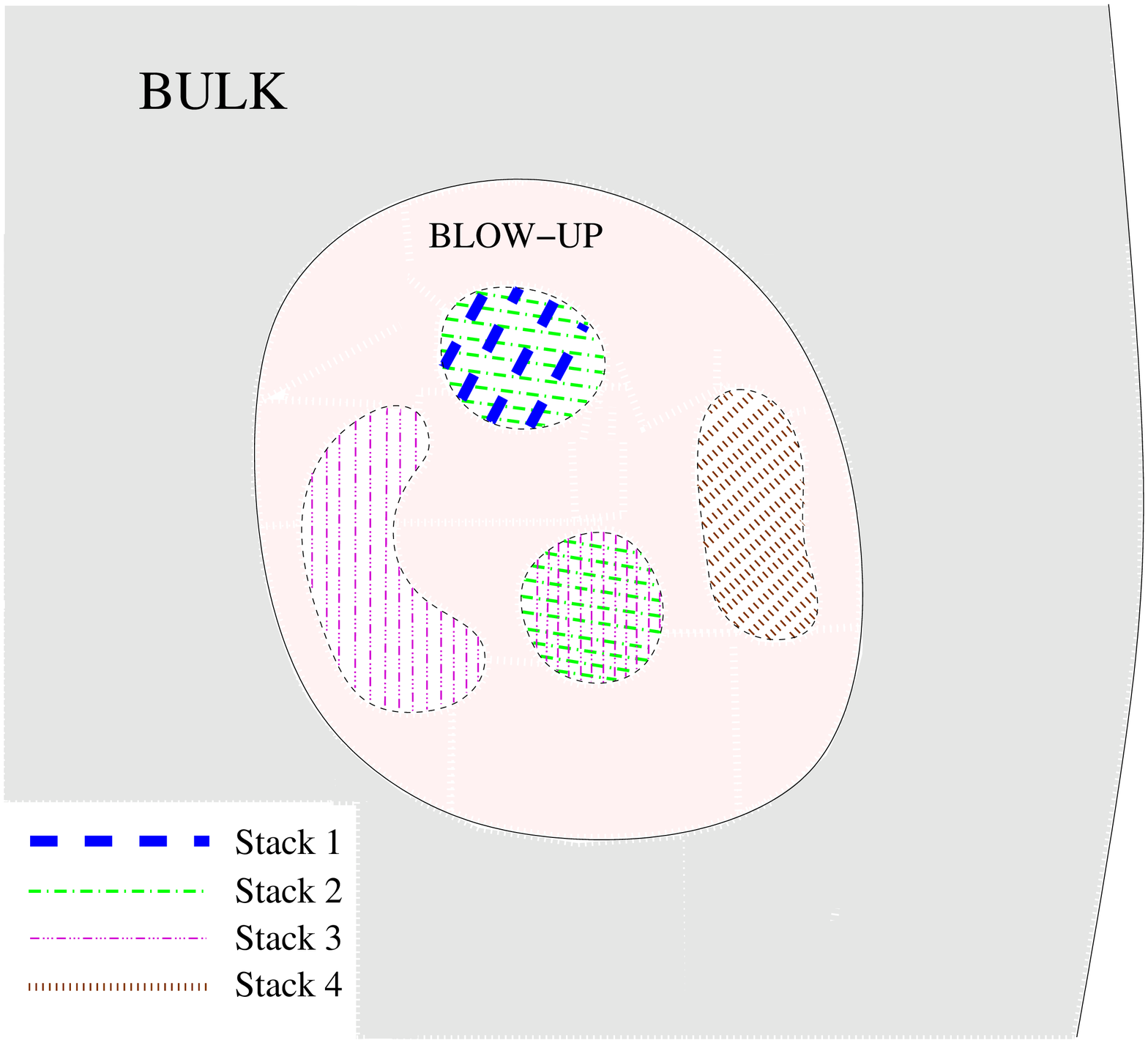}}\caption{
In the minimal geometry, there is only one small 4-cycle. The
  different brane stacks of the Standard Model are distinguished by
  having different magnetic fluxes on the internal 2-cycles of the
  4-cycle. In the minimal model above, these 2-cycles are not inherited from
  the Calabi-Yau and only exist as cycles in the geometry of the
  4-cycle. Four distinct brane stacks are required to realise the
  Standard Model, and we schematically show how these stacks are distinguished by different 
choices of magnetic flux.}\label{MinimalModel}}

\subsection{Soft Term Computations} \label{secstc} \linespread{1.2}

We now want to use the above expression (\ref{MatterMetric}) for the
matter metric in order to compute the soft terms. However, for much
of the calculations we will instead use the more general expression
(\ref{FactorisedK}). 
This will allow
greater clarity of expression and will make apparent certain
cancellations that only hold for the particular powers present in
(\ref{MatterMetric}). We will first compute the soft terms assuming
a diagonal metric and in section \ref{Nondiagonal} generalise it for the non-diagonal case.

Several results follow simply from the large-volume geometry illustrated in figure \ref{LargeVolumePic}.
We first note that in the large-volume scenario, the F-term for
the large modulus can be computed directly.
\be
F^b = e^{\hat{K}/2} \hat{K}^{b \bar{n}} \left(\partial_{\bar{n}} \bar{\hat{W}} + (\partial_{\bar{n}} \hat{K}) \bar{\hat{W}}\right).
\ee
It is a property of the K\"ahler potential $\hat{K} = - 2
\ln \mc{V}$ that
\be
\hat{K}^{m \bar{n}} \partial_{\bar{n}} \hat{K} = - 2 \tau_m.
\ee
In deriving this it is useful to recall that, as $\hat{K} = \hat{K}(T_i + \bar{T}_i)$, any function $\Omega(K)$ satisfies
\be
\partial_j \Omega(\hat{K}) \equiv \frac{\partial}{\partial T_j} \Omega(\hat{K}) =
\partial_{\bar{j}} \Omega(\hat{K}) \equiv \frac{\partial}{\partial
  \bar{T}_j} \Omega(\hat{K}) = \half \frac{\partial}{\partial \tau_j} \Omega(\hat{K}).
\ee
The stabilisation of the small moduli is such that $\partial_{\bar{i}} \bar{W}
\sim e^{-\bar{T}_i} \sim \mc{V}^{-1}$, while $\hat{K}^{b \bar{i}} \sim \mc{V}^{2/3}$.
Thus $\hat{K}^{b \bar{i}} \partial_{\bar{i}} \bar{\hat{W}} \sim \hat{K}^{b
  \bar{i}} e^{-T_i} \sim \mc{V}^{-1/3}$, and so
\be
\label{LargeFTerm}
F^b = -2 \tau_b m_{3/2} \left( 1 + \mc{O}(\mc{V}^{-1}) \right).
\ee

This implies that
any soft breaking term solely depending on $F^b$ is naturally
of order the gravitino mass $m_{3/2}$. However, it was established
 in \cite{hepth0605141}
that the F-terms of the smaller K\"ahler moduli are hierarchically smaller:
\be
F^i \sim 2 \tau^i  \frac{m_{3/2}}{\log m_{3/2}},
\ee
in units of $M_P=1$. Therefore if (as will occur)
the dependence of the soft terms on $F^b$ is cancelled, the soft terms will naturally
be smaller than the gravitino mass
by a factor of $\log(M_P/m_{3/2})\sim 30-40$.

Using (\ref{LargeFTerm}) and (\ref{FactorisedK}) the expression (\ref{SoftMassDiagFormula})
for the soft scalar masses is easily seen to become
\be
\label{SoftMassExp}
m_\alpha^2 = (1- p_\alpha)m_{3/2}^2 - \bar{F}^{\bar{i}}F^j \partial_{\bar{i}} \partial_j \log k_\alpha(\tau_i, \phi),
\ee
where as a reminder the index $i$ runs over small moduli only.

Now consider the A-terms (\ref{ATermFormula}). One simplification that can be made is that, at least in perturbation theory,
\be
\label{FY=0}
F^m \partial_m \log Y_{\alpha \beta \gamma} = 0.
\ee
The reason for this is that the superpotential does not depend on the K\"ahler moduli $T_i$.
These have perturbative shift symmetries $T \to T + i \epsilon$ that
preclude any polynomial power of $T$ appearing in the superpotential
Yukawas $Y_{\alpha \beta \gamma}$, which thus depend only on complex
structure moduli.\footnote{We neglect any nonperturbative $e^{-T}$ dependence; this is volume-suppressed and tiny.}
However, it is only the K\"ahler moduli that have non-vanishing F-terms and thus the sum in (\ref{FY=0}) vanishes.
(\ref{ATermFormula}) therefore becomes
\be
A_{\alpha \beta \gamma} = F^m \left[ \hat{K}_m - \partial_m \log (\tilde{K}_{\alpha} \tilde{K}_\beta \tilde{K}_\gamma) \right].
\ee
Now,
\bea
F^m \hat{K}_m & = & e^{\hat{K}/2} \hat{K}^{m \bar{n}} D_{\bar{n}} \bar{\hat{W}} \hat{K}_m \\
& & = e^{\hat{K}/2} \left[ \sum_n - 2 \tau_n ( \partial_{\bar{n}} \bar{\hat{W}} + (\partial_{\bar{n}} \hat{K}) \bar{\hat{W}}) \right],
\eea
where we have used $\hat{K}^{m \bar{n}} \hat{K}_{m} = - 2 \tau_n$. As $\hat{K} = - 2 \log \mc{V}$,
$\partial_{\bar{n}} \hat{K} = -2 \frac{\partial_{\bar{n}} \mc{V}}{\mc{V}}$. $\mc{V}$ is homogeneous of degree 3/2 in the
$\tau_n$, and thus
$$
\sum \tau_n \partial_{\bar{n}} \mc{V} = \frac{3 \mc{V}}{4}.
$$
Recalling that $\partial_{\bar{n}} \bar{W} \sim \mc{V}^{-1}$ for the large-volume models, we get
\be
\label{Relation1}
F^m \hat{K}_m = 3 m_{3/2} (1 + \mc{O}(\mc{V}^{-1})).
\ee
This gives
\bea
A_{\alpha \beta \gamma} & = & 3 m_{3/2} - F^m \partial_m \log (\tilde{K}_\alpha \tilde{K}_\beta \tilde{K}_\gamma) \\
& = & 3 m_{3/2} - F^b \partial_b \log(\tilde{K}_\alpha \tilde{K}_\beta \tilde{K}_\gamma) - F^{i} \partial_{i}
\log (\tilde{K}_\alpha \tilde{K}_\beta \tilde{K}_\gamma ).
\eea
Now,
$$
\log (\tilde{K}_{\alpha} \tilde{K}_{\beta} \tilde{K}_{\gamma}) = - (p_{\alpha} + p_{\beta} + p_{\gamma}) \log \tau_b
+ \log (k_{\alpha} k_{\beta} k_{\gamma}(\tau_i, \phi)),
$$
and so
\be
\partial_b \log (\tilde{K}_{\alpha} \tilde{K}_{\beta} \tilde{K}_{\gamma}) = - \frac{p_{\alpha} + p_{\beta} + p_{\gamma}}{2 \tau_b},
\ee
(where we have used $\frac{\partial}{\partial T_b} = \half \frac{\partial}{\partial \tau_b}$).
Thus we obtain
\be
\label{ATermExp}
A_{\alpha \beta \gamma} = \Big( 3 - (p_{\alpha} + p_{\beta} +
p_{\gamma})\Big) m_{3/2} - F^i \partial_i \log (k_{\alpha} k_{\beta}
k_{\gamma}).
\ee

We finally want to derive an expression for the B-term (\ref{BTermFormula}). In doing so we will use the
result of \cite{CCQPaper} that $\mu = 0$ -
this considerably simplifies expression
(\ref{BTermFormula}) as many terms vanish. We expand in full:
\bea
B \hat{\mu} & = & \frac{1}{(\tilde{K}_{H_1} \tilde{K}_{H_2})^\half} \Bigg[ (2 m_{3/2}^2 + V_0)Z \nonumber \\
& & + m_{3/2} \left( F^m \partial_m Z - F^{\bar{m}} \partial_{\bar{m}} Z \right) - m_{3/2} Z F^m \partial_m \log
\left(\tilde{K}_{H_1} \tilde{K}_{H_2} \right) - \nonumber \\
& & \bar{F}^{\bar{m}} F^n \left[ \partial_{\bar{m}} \partial_n Z -
(\partial_{\bar{m}} Z) \partial_n \log \left( \tilde{K}_{H_1} \tilde{K}_{H_2} \right) \right] \Bigg] \\
& = &  \frac{1}{(\tilde{K}_{H_1} \tilde{K}_{H_2})^\half} \Bigg[(2 m_{3/2}^2 + V_0)Z + m_{3/2} \left( F^m \partial_m Z
- F^{\bar{m}} \partial_{\bar{m}} Z \right) \nonumber \\
& & - m_{3/2} Z \left[
F^b \partial_b \log(\tilde{K}_{H_1} \tilde{K}_{H_2}) + F^i \partial_i \log \left( \tilde{K}_{H_1} \tilde{K}_{H_2} \right) \right] \nonumber \\
& & - \bar{F}^{\bar{b}} F^b \left[ \partial_{\bar{b}} \partial_b Z - (\partial_{\bar{b}} Z) \partial_b
\log (\tilde{K}_{H_1} \tilde{K}_{H_2})\right] \nonumber \\
& &  - \bar{F}^{\bar{b}} F^i \left[ \partial_{\bar{b}} \partial_i Z - (\partial_{\bar{b}} Z) \partial_i
\log (\tilde{K}_{H_1} \tilde{K}_{H_2})\right] \nonumber \\
& &  - \bar{F}^{\bar{j}} F^b \left[ \partial_{\bar{j}} \partial_b Z - (\partial_{\bar{j}} Z) \partial_b
\log (\tilde{K}_{H_1} \tilde{K}_{H_2})\right] \nonumber \\
& &  - \bar{F}^{\bar{j}} F^i \left[ \partial_{\bar{j}} \partial_i Z - (\partial_{\bar{j}} Z) \partial_i
\log (\tilde{K}_{H_1} \tilde{K}_{H_2})\right] \Bigg],
\eea
which gives
\bea
B \hat{\mu} & = & \frac{\tau_b^{\frac{p_1+p_2}{2}}}{(k_{H_1} k_{H_2})^{\half}} \Bigg[ (2 m_{3/2}^2 + V_0) Z
+ \tau_b^{-p_z} m_{3/2} \left(F^i \partial_i z - F^{\bar{i}} \partial_{\bar{i}} z \right) \nonumber \\
& & - m_{3/2} Z \left[-2 \tau_b m_{3/2} \partial_b \Big(-(p_1 + p_2) \log \tau_b \Big) + F^i \partial_i \log (k_{H_1} k_{H_2}) \right] \nonumber \\
& & -(2 \tau_b)^2 m_{3/2}^2 \left[ \partial_{\bar{b}} \partial_b (\tau_b^{-p_z} z) - \partial_{\bar{b}} (\tau_b^{-p_z} z)
\partial_b \Big(-(p_1 + p_2) \log \tau_b \Big) \right] \nonumber \\
& & + (2 \tau_b) m_{3/2} F^i \left[  \partial_{\bar{b}} \partial_i (\tau_b^{-p_z} z) - \partial_{\bar{b}} (\tau_b^{-p_z} z)
\partial_i \log (k_{H_1} k_{H_2}) \right] \nonumber \\
& & + (2 \tau_b) m_{3/2} \bar{F}^{\bar{j}} \left[  \partial_{\bar{j}} \partial_b (\tau_b^{-p_z} z) - \partial_{\bar{j}} (\tau_b^{-p_z} z)
\partial_b \log  \Big( -(p_1 + p_2) \log \tau_b \Big) \right] \nonumber \\
& & - \bar{F}^{\bar{j}} F^i \left[ \partial_{\bar{j}} \partial_i Z - (\partial_{\bar{j}} Z) \partial_i \log (\tilde{K}_{H_1}
\tilde{K}_{H_2}) \right] \Bigg].
\eea
This becomes
\bea
B \hat{\mu} & = & \frac{\tau_b^{\frac{p_1 + p_2}{2} - p_z}}{(k_{H_1} k_{H_2})^{\half}} \Bigg[ \Big(2 - (p_1 + p_2)
+ p_z(p_1 + p_2) - p_z(p_z + 1)\Big) m_{3/2}^2 z + V_0 z  \nonumber \\
& & + m_{3/2} \left( F^i \partial_i z - F^{\bar{i}} \partial_{\bar{i}} z \right)
- m_{3/2} z F^i \partial_i \log (k_{H_1} k_{H_2}) \nonumber \\
& & - m_{3/2} p_z (F^i \partial_i z + \bar{F}^{\bar{j}}\partial_{\bar{j}} z )
+p_z z F^i \partial_i \log (k_{H_1} k_{H_2}) + (p_1 + p_2) \bar{F}^{\bar{j}} \partial_{\bar{j}} z \Bigg] \nonumber \\
& & + \frac{\tau_b^{\frac{p_1 + p_2}{2}}}{(k_{H_1} k_{H_2})^{\half}} \left[ - \bar{F}^{\bar{j}} F^i \left[
\partial_{\bar{j}} \partial_i Z - (\partial_{\bar{j}} Z) \partial_i \log \left( \tilde{K}_{H_1} \tilde{K}_{H_2} \right) \right] \right],
\eea
giving
\bea
\label{BMuExpression}
B \hat{\mu} & = & \frac{\tau_b^{\frac{p_1 + p_2}{2} - p_z}}{(k_{H_1} k_{H_2})^{\half}} \Bigg[
\Big( 2 - (p_1 + p_2)(1-p_z) - p_z(p_z+1) \Big) m_{3/2}^2 z + V_0 z \nonumber \\
& & + (p_z - 1)z F^i \partial_i  \log (k_{H_1} k_{H_2} ) m_{3/2}
+ m_{3/2} \left( (1-p_z) F^i \partial_i z + (p_1 + p_2 -p_z -1)F^{\bar{j}}\partial_{\bar{j}}z \right)\Bigg] \nonumber \\
& & -  \frac{\tau_b^{\frac{p_1 + p_2}{2} - p_z}}{(k_{H_1} k_{H_2})^{\half}} \Bigg[
\bar{F}^{\bar{j}} F^i \left[ \partial_{\bar{j}} \partial_i z - \partial_{\bar{j}} z \partial_i \log (k_{H_1} k_{H_2}) \right]
\Bigg].
\eea

We now note that when $p_{\alpha} =1$, there are cancellations in each of equations (\ref{SoftMassExp}), (\ref{ATermExp}) and
(\ref{BMuExpression}). This is an independent illustration of the the
critical nature of the
value $p_{\alpha} = 1$ that appears in equation (\ref{MatterMetric}).

If we now use the correct value $p_\alpha = 1$, and also assume a vanishing cosmological constant $V_0 =0$, we obtain
\bea
\hat{Y}_{\alpha \beta \gamma} & = & \frac{Y_{\alpha \beta \gamma}}{(k_{\alpha} k_{\beta} k_{\gamma})^{\half}}, \\
\hat{\mu} & = & - \frac{F^{\bar{i}} \partial_i z}{(k_{H_1} k_{H_2})^{\half}}, \\
M_i & = & \frac{F^i}{2 \tau_i}, \\
m_\alpha^2 & = & - \sum_{i, \bar{j}} F^i \bar{F}^{\bar{j}} \partial_i \partial_{\bar{j}} \log k_\alpha(\tau_i), \\
A_{\alpha \beta \gamma} & = & - \sum_i F^i \partial_i \log (k_\alpha k_\beta k_\gamma), \\
B \hat{\mu} & = & - \frac{1}{(k_{H_1} k_{H_2})^{\half}} \left[ \bar{F}^{\bar{j}} F^i \left[ \partial_{\bar{j}} \partial_i z
- \partial_{\bar{j}} z \partial_i \log (\tilde{k}_{H_1} \tilde{k}_{H_2}) \right] \right].
\eea

We further simplify these expressions by expanding
$k_\alpha(\tau_s, \phi)$ and $z(\tau_s, \phi)$ as a power series in $\tau_s$,
\bea
k_{\alpha}(\tau_s, \phi) & = & \tau_s^{\lambda} k^0_{\alpha}(\phi) +
\mc{O}(\tau_s^{\lambda -1})k^1_{\alpha} (\phi) + \ldots, \\
z(\tau_s, \phi) & = & \tau_s^{\lambda} z^0(\phi) +  \mc{O}(\tau_s^{\lambda - 1})z^1(\phi) + \ldots.
\eea
We assume $k_{\alpha}$ and $z$ have the same scaling power $\lambda$.
The soft parameters then become
\bea
\label{YukMin}
\hat{Y}_{\alpha \beta \gamma} & = & \frac{Y_{\alpha \beta \gamma}}{ \tau_s^{\frac{3 \lambda}{2}}
(k^0_{\alpha} k^0_{\beta} k^0_{\gamma}(\phi))^{\half}}, \\
\hat{\mu} & = & - \left( \frac{F^{\bar{s}}}{2 \tau_s} \right)
\lambda \frac{z^0(\phi)}
{ (k^0_{H_1} k^0_{H_2})^{\half}}, \\
M_i & = & \frac{F^s}{2 \tau_s}, \\
m_\alpha^2 & = & \lambda \left(\frac{F^s}{2 \tau_s}\right) \left( \frac{\bar{F}^{\bar{s}}}{2 \tau_s} \right),\\
A_{\alpha \beta \gamma} & = & - 3 \lambda \left(\frac{F^s}{2 \tau_s}\right) \\
\label{BMin}
B \hat{\mu} & = & \frac{ z^0(\phi)}
{(k^0_{H_1} k^0_{H_2}(\phi))^{\half}} \lambda(\lambda + 1) \left( \frac{F^s}{2 \tau_s} \right)
\left( \frac{\bar{F}^{\bar{s}}}{2 \tau_s} \right) \equiv -\left( \frac{F^s}{2 \tau_s} \right) (\lambda + 1) \hat{\mu}.
\eea
The expressions (\ref{YukMin}) to (\ref{BMin}) are appealingly simple. Note that the supersymmetric parameters, namely
$Y_{\alpha \beta \gamma}$ and $\mu$, both require knowledge of the flavour sector through the
complex structure moduli.
However the soft parameters are entirely set by $\lambda$ and $F^s$.
As discussed in \cite{CCQPaper}, $0 < \lambda < 1$, and for the geometry of the minimal model
$\lambda = 1/3$.
In this case, the pure soft parameters
in the dilute flux approximation become
\bea
\label{eq1}
M_i & = & \frac{F^s}{2 \tau_s}, \\
\label{eq2}
m_{\alpha} & = & \frac{1}{\sqrt{3}} M_i, \\
\label{eq3}
A_{\alpha \beta \gamma} & = & - M_i, \\
\label{eq4}
B & = & - \frac{4}{3} M_i.
\eea
It is amusing to note that the scalars, gauginos and A-terms in
expressions (\ref{eq1}) to (\ref{eq4}) are identical to that of the dilaton-dominated scenario that was
much studied in heterotic models.
Notice that all soft terms are proportional to $ \frac{F^s}{2
  \tau_s}\sim m_{3/2}/\log(M_P/ m_{3/2})$
and are therefore reduced with respect to the gravitino
mass.

\subsection{General Non-Diagonal Matter Metrics}
\label{Nondiagonal}

We now want to extend the above formulae to the case
of arbitrary non-diagonal matter metrics. The expression for the normalised
gaugino masses is unaltered
\be
M_i = \frac{F^i}{2 \tau_i}.
\ee
The soft scalar Lagrangian is
\be
\mc{L}_{soft} = \tilde{K}_{\alpha \bar{\beta}} \partial_\mu C^\alpha
\partial^\mu \bar{C}^{\bar{\beta}} - \tilde{m}^{2}_{\alpha \bar{\beta}} C^{\alpha}
\bar{C}^{\bar{\beta}} - \left( \frac{1}{6} A'_{\alpha \beta \gamma} C^{\alpha} C^{\beta} C^{\gamma} +
B\hat{\mu}H_1 H_2 +  c.c \right),
\ee
where \cite{bim}
\bea
\label{GenSoftMass}
\tilde{m}_{\alpha \bar{\beta}}^{2} & = & (m_{3/2}^2 + V_0) \tilde{K}_{\alpha
  \bar{\beta}}
- \bar{F}^{\bar{m}} F^n \left( \partial_{\bar{m}} \partial_n \tilde{K}_{\alpha \bar{\beta}}
- (\partial_{\bar{m}} \tilde{K}_{\alpha \bar{\gamma}}) \tilde{K}^{\bar{\gamma} \delta} (\partial_n \tilde{K}_{\delta \bar{\beta}})
\right), \\
\label{GenAExp}
A'_{\alpha \beta \gamma} & = & e^{\hat{K}/2} F^m \Big[ \hat{K}_m Y_{\alpha \beta \gamma} +
\partial_m  Y_{\alpha \beta \gamma} \nonumber \\
& & - \left( (\partial_m \tilde{K}_{\alpha \bar{\rho}})
\tilde{K}^{\bar{\rho} \delta} Y_{\delta \beta \gamma} + (\alpha
\leftrightarrow \beta) + (\alpha \leftrightarrow \gamma) \right)
\Big].
\eea

As with the diagonal case, we can extract the overall volume dependence to write
\be
\tilde{K}_{\alpha \bar{\beta}} = \tau_b^{-p} k_{\alpha \bar{\beta}}(\tau_i, \phi_i).
\ee
Again $p=1$, but as above
we for now leave $p$ unspecified.
We first compute the Lagragian parameters $\tilde{m}^{2}_{\alpha \bar{\beta}}$ and $A'_{\alpha \beta \gamma}$,
before subsequently considering the physical masses and couplings.

The expression for soft masses becomes
\bea
\tilde{m}_{\alpha \bar{\beta}}^{2} & = & (m_{3/2}^2 + V_0) \tilde{K}_{\alpha \bar{\beta}} \nonumber \\
& & - \bar{F}^{\bar{b}} F^b \left[ \partial_{\bar{b}} \partial_b  \left( \tau_b^{-p} k_{\alpha \bar{\beta}} \right)
- \partial_{\bar{b}} \left( \tau_b^{-p} k_{\alpha \bar{\gamma}} \right) \left( \tau_b^p k^{\bar{\gamma} \delta} \right)
\partial_b \left( \tau_b^{-p} k_{\delta \bar{\beta}} \right) \right] \nonumber \\
& &  - \bar{F}^{\bar{b}} F^i \left[ \partial_{\bar{b}} \partial_i  \left( \tau_b^{-p} k_{\alpha \bar{\beta}} \right)
- \partial_{\bar{b}} \left( \tau_b^{-p} k_{\alpha \bar{\gamma}} \right) \left( \tau_b^p k^{\bar{\gamma} \delta} \right)
\partial_i \left( \tau_b^{-p} k_{\delta \bar{\beta}} \right) \right] \nonumber \\
& &  - \bar{F}^{\bar{i}} F^b \left[ \partial_{\bar{i}} \partial_b  \left( \tau_b^{-p} k_{\alpha \bar{\beta}} \right)
- \partial_{\bar{i}} \left( \tau_b^{-p} k_{\alpha \bar{\gamma}} \right) \left( \tau_b^p k^{\bar{\gamma} \delta} \right)
\partial_b \left( \tau_b^{-p} k_{\delta \bar{\beta}} \right) \right] \nonumber \\
& &  - \bar{F}^{\bar{i}} F^j \left[ \partial_{\bar{i}} \partial_j  \left( \tau_b^{-p} k_{\alpha \bar{\beta}} \right)
- \partial_{\bar{i}} \left( \tau_b^{-p} k_{\alpha \bar{\gamma}} \right) \left( \tau_b^p k^{\bar{\gamma} \delta} \right)
\partial_j \left( \tau_b^{-p} k_{\delta \bar{\beta}} \right) \right].
\eea
Putting $V_0 = 0$, this becomes
\bea
\tilde{m}_{\alpha \bar{\beta}}^{2} & = & m_{3/2}^2 \tau_b^{-p} k_{\alpha \bar{\beta}}(\tau_i) \nonumber \\
& & - (2 \tau_b)^2 m_{3/2}^2 \left[ \frac{p(p+1)}{2 \tau_b} \left( \tau_b^{-p} k_{\alpha \bar{\beta}} \right)
- \left( \frac{(-p)}{2 \tau_b} k_{\alpha \bar{\gamma}} \tau_b^{-p} (\tau_b^p
k^{\bar{\gamma} \delta} ) \frac{(-p)}{2 \tau_b} \tau_b^{-p} k_{\delta \bar{\beta}} \right) \right] \nonumber \\
& & - (2 \tau_b) m_{3/2} F^i \left[ \frac{(-p)}{2 \tau_b} \tau_b^{-p} \partial_i k_{\alpha \bar{\beta}} -
\left( \frac{(-p)}{2 \tau_b} k_{\alpha \bar{\gamma}} \tau_b^{-p} (\tau_b^p k^{\bar{\gamma} \delta}) \tau_b^{-p} \partial_i
k_{\delta \bar{\beta}} \right) \right] \nonumber \\
& & - (2 \tau_b) m_{3/2} F^{\bar{j}} \left[ \frac{(-p)}{2 \tau_b} \tau_b^{-p} \partial_{\bar{j}} k_{\alpha \bar{\beta}} -
\left( \frac{(-p)}{2 \tau_b} k_{\alpha \bar{\gamma}} \tau_b^{-p} (\tau_b^p k^{\bar{\gamma} \delta}) \tau_b^{-p} \partial_{\bar{j}}
k_{\delta \bar{\beta}} \right) \right] \nonumber \\
& & - F^i \bar{F}^{\bar{j}} \left[ \tau_b^{-p} \partial_i \partial_{\bar{j}} k_{\alpha \bar{\beta}} -
\tau_b^{-p} (\partial_i k_{\alpha \bar{\gamma}} )(k^{\bar{\gamma} \delta} \tau_b^p)(\tau_b^{-p} \partial_{\bar{j}}
k_{\delta \bar{\beta}}) \right].
\eea
Contracting indices we obtain
\bea
\tilde{m}_{\alpha \bar{\beta}}^{2} & = & m_{3/2}^2 (\tau_b^{-p} k_{\alpha \bar{\beta}}) - m_{3/2}^2 \left( p(p+1) - p^2 \right)
(k_{\alpha \bar{\beta}} \tau_b^{-p}) \nonumber \\
& & - m_{3/2} F^i \tau_b^{-p} \left( -p \partial_i k_{\alpha \bar{\beta}} + p \partial_i k_{\alpha \bar{\beta}} \right)
- m_{3/2} \bar{F}^{\bar{j}} \tau_b^{-p} \left(-p \partial_{\bar{j}} k_{\alpha \bar{\beta}} +
p \partial_{\bar{j}} k_{\alpha \bar{\beta}} \right) \nonumber \\
& & - F^i F^{\bar{j}} \tau_b^{-p} \left( \partial_i \partial_j k_{\alpha \bar{\beta}} -
(\partial_i k_{\alpha \bar{\gamma}}) k^{\bar{\gamma} \delta} (\partial_{\bar{j}} k_{\delta \bar{\beta}}) \right).
\eea
This simplifies to
\be
\label{SoftMasses2}
\tilde{m}_{\alpha \bar{\beta}}^{2} =  \tau_b^{-p} m_{3/2}^2 k_{\alpha \bar{\beta}} (1-p)
- F^i \bar{F}^{\bar{j}} \tau_b^{-p} \left( \partial_i \partial_{\bar{j}} k_{\alpha \bar{\beta}}
- (\partial_i k_{\alpha \bar{\gamma}}) k^{\bar{\gamma} \delta} (\partial_{\bar{j}} k_{\delta \bar{\beta}}) \right).
\ee
For the critical value of $p=1$, (\ref{SoftMasses2}) simplifies to
\be
m_{\alpha \bar{\beta}}^{'2} = - \frac{F^i \bar{F}^{\bar{j}}}{\tau_b} \left( \partial_i \partial_{\bar{j}} k_{\alpha \bar{\beta}}
- (\partial_i k_{\alpha \bar{\gamma}}) k^{\bar{\gamma} \delta} (\partial_{\bar{j}} k_{\delta \bar{\beta}}) \right).
\ee

Now consider the A-terms (\ref{GenAExp}).
As before, the K\"ahler moduli do not appear
in the superpotential and thus
\be
F^m \partial_m Y_{\alpha \beta \gamma} = 0.
\ee
This simplifies (\ref{GenAExp}) to
\be
A'_{\alpha \beta \gamma} = e^{\hat{K}/2} F^m \left[ \hat{K}_m Y_{\alpha \beta \gamma} - \left(
(\partial_m \tilde{K}_{\alpha \bar{\rho}}) \tilde{K}^{\bar{\rho} \delta} Y_{\delta \beta \gamma} +
(\alpha \leftrightarrow \beta) + (\alpha \leftrightarrow \gamma) \right) \right].
\ee
The calculation of $F^m \hat{K}_m$ depends only on the moduli fields and thus the previous result (\ref{Relation1}) is unaltered,
\be
F^m \hat{K}_m = 3 m_{3/2} \left( 1 + \mc{O}(\mc{V}^{-1}) \right),
\ee
giving
\be
\label{AExp2}
A'_{\alpha \beta \gamma} = e^{\hat{K}/2} \left[ 3 m_{3/2} Y_{\alpha \beta \gamma} -
F^m \left((\partial_m \tilde{K}_{\alpha \bar{\rho}} ) \tilde{K}^{\bar{\rho} \delta} Y_{\delta \beta \gamma}
+ (\alpha \leftrightarrow \beta) + (\alpha \leftrightarrow \gamma) \right) \right].
\ee
We split the sum over moduli in (\ref{AExp2}),
\bea
A'_{\alpha \beta \gamma} & = & e^{\hat{K}/2} \Big[  3 m_{3/2} Y_{\alpha \beta \gamma} \nonumber \\
& & - F^b \left((\partial_b \tilde{K}_{\alpha \bar{\rho}} ) \tilde{K}^{\bar{\rho} \delta} Y_{\delta \beta \gamma}
+ (\alpha \leftrightarrow \beta) + (\alpha \leftrightarrow \gamma) \right) \Big] \nonumber \\
& & - F^i \left((\partial_i \tilde{K}_{\alpha \bar{\rho}} ) \tilde{K}^{\bar{\rho} \delta} Y_{\delta \beta \gamma}
+ (\alpha \leftrightarrow \beta) + (\alpha \leftrightarrow \gamma) \right) \\
& = &  e^{\hat{K}/2} \Big[  3 m_{3/2} Y_{\alpha \beta \gamma} \nonumber \\
& & -  (-2 \tau_b m_{3/2}) \left( \frac{(-p)}{2 \tau_b} \tau_b^{-p} k_{\alpha \bar{\rho}}
(\tau_b^p k^{\bar{\rho} \delta}) Y_{\delta \beta \gamma} + (\alpha \leftrightarrow \beta) + (\alpha \leftrightarrow \gamma) \right) \nonumber \\
& & - F^i \left( (\partial_i \tilde{K}_{\alpha \bar{\rho}}) \tilde{K}^{\bar{\rho} \delta} Y_{\delta \beta \gamma} +
(\alpha \leftrightarrow \beta) + (\alpha \leftrightarrow \gamma) \right) \Big].
\label{ATerm2}
\eea
Simplifying (\ref{ATerm2}) we obtain
\be
A_{\alpha \beta \gamma} = e^{\hat{K}/2} \left[ 3(1-p) m_{3/2} Y_{\alpha \beta \gamma} -
F^i \left( (\partial_i k_{\alpha \bar{\rho}}) k^{\bar{\rho} \delta} Y_{\delta \beta \gamma}
+ (\alpha \leftrightarrow \beta) + (\alpha \leftrightarrow \gamma) \right) \right].
\ee
Imposing the critical value $p=1$, we get
\be
A'_{\alpha \beta \gamma} =  - e^{\hat{K}/2} F^i \left( (\partial_i k_{\alpha \bar{\rho}}) k^{\bar{\rho} \delta} Y_{\delta \beta \gamma}
+ (\alpha \leftrightarrow \beta) + (\alpha \leftrightarrow \gamma) \right).
\ee

There is no family replication in the Higgs sector, and so for the $\mu$ and $B$ terms the
issues of non-diagonal metrics do not apply. As with the diagonal case,
the unnormalised physical $\mu$ term is
\be
\mu' = e^{\hat{K}/2} \mu + m_{3/2} (1-p_z) z - (F^{\bar{i}} \partial_{\bar{i}} z) \tau_b^{-p_z}.
\ee
Imposing $p=1$ and the vanishing of the superpotential $\mu$ term, we obtain
\be
\mu' = - \frac{F^{\bar{i}} \partial_{\bar{i}} z}{\tau_b}.
\ee
The normalised $\mu$ term remains
\be
\hat{\mu} = - \frac{F^{\bar{i}} \partial_i z}{(k_{H_1} k_{H_2}(\tau_i, \phi))^{\half}}.
\ee
The B-term calculation is likewise unaltered and
the general un-normalised B term can be read off from (\ref{BMuExpression}). Putting $p=1$, we again obtain
\be
B \hat{\mu} = - F^{\bar{j}} F^i \left[ \partial_{\bar{j}} \partial_i z - (\partial_{\bar{j}} z) \partial_i
\log \left( k_{H_1} k_{H_2} \right) \right].
\ee
where we have assumed $p_z = p_1 = p_2 = 1$.
Finally, the unnormalised Yukawa couplings are
\be
Y'_{\alpha \beta \gamma} = e^{\hat{K}/2} Y_{\alpha \beta \gamma}.
\ee

Gathering together the results on soft terms, we have
\bea
\tilde{K}_{\alpha \bar{\beta}} & = & \frac{k_{\alpha \bar{\beta}}(\tau_i,\phi)}{\tau_b}, \\
Y'_{\alpha \beta \gamma} & = & e^{\hat{K}/2} Y_{\alpha \beta \gamma}, \\
\mu' & = & - (F^{\bar{i}} \partial_{\bar{i}} z) \tau_b^{-p_z}, \\
M_i & = & \frac{F^i}{2 \tau_i}, \\
m_{\alpha \bar{\beta}}^{'2} & = & - \frac{F^i \bar{F}^{\bar{j}}}{\tau_b} \left( \partial_i \partial_{\bar{j}} k_{\alpha \bar{\beta}}
- (\partial_i k_{\alpha \bar{\gamma}}) k^{\bar{\gamma} \delta} (\partial_{\bar{j}} k_{\delta \bar{\beta}}) \right), \\
A'_{\alpha \beta \gamma} & = & - e^{\hat{K}/2} F^i \left( (\partial_i k_{\alpha \bar{\rho}}) k^{\bar{\rho} \delta} Y_{\delta \beta \gamma}
+ (\alpha \leftrightarrow \beta) + (\alpha \leftrightarrow \gamma) \right), \\
B \hat{\mu} & = & - F^{\bar{j}} F^i \left[ \partial_{\bar{j}} \partial_i z - (\partial_{\bar{j}} z) \partial_i
\log \left( k_{H_1} k_{H_2} \right) \right].
\eea
Notice again that having imposed $p=1$
the F-term of the large modulus $F^b$ does not appear in these expressions  and that all the soft-terms are proportional to
$ \frac{F^i}{2 \tau_i}\sim m_{3/2}/\log(m_{3/2})$ and are therefore reduced with respect to the gravitino
mass.

Following the same calculational steps as for the diagonal case we now perform a power series expansion of
$k_{\alpha \bar{\beta}}$ in terms of the small modulus $\tau_s$.
We write
\bea
\label{MatterMetricND}
k_{\alpha \bar{\beta}}(\tau_s, \phi_j) & = & \tau_s^{\lambda} \tilde{k}_{\alpha \bar{\beta}}(\phi)
+ \mc{O}(\tau_s^{\lambda -1}) \tilde{k}^1_{\alpha \bar{\beta}}(\phi) +
\ldots, \nonumber \\
\tilde{K}_{\alpha \bar{\beta}} & = & \frac{\tau_s^{\lambda} \tilde{k}_{\alpha \bar{\beta}}(\phi)}{\tau_b}.
\eea
As discussed further in \cite{CCQPaper} the subleading terms in (\ref{MatterMetricND}) should be interpreted as loop
corrections. For the geometry of the minimal model, $\lambda = 1/3$.

As discussed in \cite{CCQPaper} the power $\lambda$ is flavour-universal.
This is because flavour is defined through the superpotential
and in particular through the Yukawa couplings. It is the Yukawa couplings, and only the Yukawa couplings,
that distinguish the top quark from the up quark.
These only depend on the complex structure moduli, and thus it is the values
of the complex structure moduli - not the K\"ahler moduli - that distinguish the different flavours.
By neglecting the complex structure moduli and focusing only
on the K\"ahler moduli dependence through the power $\lambda$, we are insensitive to flavour.
$\lambda$ will thus be
universal for fields with different flavours but the same gauge charges.
By contrast, the function $k_{\alpha \bar{\beta}}$ is flavour-dependent,
as it is sensitive to the complex structure moduli which source flavour.

However, the form (\ref{MatterMetricND}) of the matter metric in fact gives soft term universality! To see this, note that
the soft terms $\tilde{m}^{2}_{\alpha \bar{\beta}}$ and $A'_{\alpha \beta \gamma}$ become
\bea
\label{softtermAm}
A'_{\alpha \beta \gamma} & = & - 3 \lambda \left( \frac{F^s}{2 \tau_s} \right) e^{\hat{K}/2} Y_{\alpha \beta \gamma}, \\
m^{'2}_{\alpha \bar{\beta}} & = & - \frac{1}{\tau_b} \frac{F^s F^{\bar{s}}}{(2 \tau_s)^2}
 \left(\lambda (\lambda - 1) - \lambda^2 \right) k_{\alpha \bar{\beta}}, \nonumber \\
& = &  \lambda \left(  \frac{ F^s F^{\bar{s}}}{(2 \tau_s)^2} \right) \tilde{K}_{\alpha \bar{\beta}}.
\eea
As the unnormalised soft scalar masses are proportional to the matter metric,
the physical masses are universal and flavour-independent. Furthermore the A-terms are also
proportional to the Yukawa couplings, and thus the scalar superpartners are diagonalised by the same transformation that
diagonalises the Standard Model fermions.
Thus the physical soft terms in the non-diagonal case are in fact identical to those found
for the diagonal case in equations (\ref{YukMin}) to (\ref{BMin}).
For the minimal model with $\lambda = 1/3$, this again gives
\bea
\label{ndeq1}
M_i & = & \frac{F^s}{2 \tau_s}, \\
\label{ndeq2}
m_{\alpha} & = & \frac{1}{\sqrt{3}} M_i, \\
\label{ndeq3}
A_{\alpha \beta \gamma} & = & - M_i, \\
\label{ndeq4}
B & = & - \frac{4}{3} M_i.
\eea
In the next section we discuss universality and explain on a more
conceptual level why the above computations led to universal soft terms.

\subsection{Universality and Gravity Mediation}
\label{Universality}

Gravity mediation is commonly held to have a generic problem with non-universality, being expected to
lead to non-diagonal
squark masses and large unobserved flavour-changing neutral currents. The intuition for this is that
flavour physics is Planck-scale physics, sensitive to the details of the compactification. In gravity mediation,
supersymmetry breaking is also Planck scale physics, and thus we expect the physics that breaks supersymmetry to
also be sensitive to flavour, thus generating non-universality.
This has encouraged the development of other approaches to supersymmetry
breaking, such as gauge \cite{gaugemed} or anomaly mediation \cite{anomalymed, randallsundrum}.
These have in common the property that the supersymmetry breaking sector is decoupled or sequestered from the
flavour sector.

A general pre-requisite for universality is that supersymmetry breaking does not see flavour.
With this in mind, let us reconsider universality in gravity-mediated models. Gravity mediation is characterised by a hidden sector
that only couples to visible matter through non-renormalisable operators.
For supersymmetry breaking to be flavour-universal,
it is necessary to be able to divide the 
hidden sector (i.e. the moduli) into (at least) two
decoupled sectors
\be
\Phi = \Psi_{\hbox{susy-breaking}} \oplus \chi_{\hbox{flavour}}.
\ee
One sector ($\Psi$) is responsible for breaking supersymmetry, while
the other ($\chi$) is responsible for breaking universality and generating the flavour structure through the Yukawa couplings.
As the $\Psi$ sector breaks supersymmetry and generates soft terms, we have
\bea
\label{psisb}
D_{\Psi} W & \neq & 0, \nonumber \\
F^{\Psi} & \neq & 0.
\eea
As the $\chi$ sector breaks universality and generates flavour, in order for soft terms to be flavour-universal
the $\chi$ sector must not break supersymmetry. We require
\bea
\label{chisb}
D_{\chi} W & = & 0, \nonumber \\
F^{\chi} & = & 0.
\eea
As $F^{\chi}$ contains a term $e^{K/2} K^{\chi \bar{\Psi}} D_{\bar{\Psi}} W$ and $D_{\bar{\Psi}} W \neq 0$,
for these equations to be satisfied we require
$K^{\chi \bar{\Psi}} = 0$. That is, the two sectors $\Psi$ and $\chi$ must be entirely
decoupled and the hidden sector metric must be block-diagonal,
\be
K_{\Phi \bar{\Phi}} = \left( \begin{array}{cc} K_{\Psi \bar{\Psi}} & 0 \\ 0 & K_{\chi \bar{\chi}}
\end{array} \right).
\ee
We suppose such a decoupling can be achieved and consider, within supergravity, the dependence of the superpotential
and K\"ahler potential on $\Psi$ and $\chi$. As $\Psi$ does not see flavour, it must be absent from the superpotential
Yukawa couplings,
\be
Y_{\alpha \beta \gamma}(\chi, \Psi) = Y_{\alpha \beta \gamma}(\chi).
\ee
Furthermore, as $\Psi$ does not see flavour the matter K\"ahler metric
can only
depend on $\Psi$ as an overall, flavour-universal,
prefactor,
\be
\label{UniMetric}
\tilde{K}_{\alpha \bar{\beta}}(\chi, \Psi) = f(\Psi) \tilde{k}_{\alpha \bar{\beta}}(\chi).
\ee
One may confirm through the formulae (\ref{GenSoftMass}) and (\ref{GenAExp})
that the above metric and superpotential with the supersymmetry
breaking of (\ref{psisb}) and (\ref{chisb}) do indeed give
physical soft terms that are flavour universal.

In the context of effective field theory, the above decoupling is obviously unnatural.
There is no reason for certain moduli to be excluded from the superpotential, and in any case
a generic hidden sector cannot be written
as a direct sum of two distinct sectors. However, this is not the case in string theory.
In string compactifications the moduli can, at leading order, be separated
into distinct sectors,
\be
\Phi = (\hbox{dilaton}) \oplus (\hbox{complex structure moduli}) \oplus (\hbox{K\"ahler moduli}).
\ee
The leading order K\"ahler potential is\footnote{for IIB compactifications - very similar expressions hold for other cases.}
\be
\hat{K} = - \ln (S + \bar{S}) - \ln \left( i \int \Omega \wedge \bar{\Omega} \right) - 2 \ln (\mc{V}),
\ee
and generates a block-diagonal metric,
\be
\label{bdmetric}
\hat{K}_{\Phi \bar{\Phi}} = \left( \begin{array}{ccc} \hat{K}_{S \bar{S}} & 0 & 0 \\
0 & \hat{K}_{U_i \bar{U}_j} & 0 \\ 0 & 0 & \hat{K}_{T_i \bar{T}_j} \end{array}  \right).
\ee
Furthermore, it is well-known that brane worldvolume theories do
exhibit a decoupling of K\"ahler and complex structure moduli. For B-type branes, the
superpotential describing the theory on the brane worldvolume depends only on complex
structure moduli, whereas for A-type branes the superpotential depends only on K\"ahler moduli \cite{hepth9906200}.
The genericity arguments for flavour non-universality in gravity-mediated supersymmetry breaking therefore do not
necessarily apply to string compactifications, where the moduli do separate into distinct sectors.
Flavour universality can be achieved if one of these sectors - the K\"ahler moduli - break supersymmetry
while a second sector - the complex structure moduli - source flavour.
 The complex structure and K\"ahler
sectors are separate but related: the symmetry that relates the sectors
but also keeps them distinct is mirror symmetry.

Our contention is that is exactly what occurs
for IIB flux models. As argued above, the constraints of holomorphy and shift symmetry
 imply that the K\"ahler moduli cannot
appear in the superpotential and do not see flavour.
The complex structure moduli do appear in
$Y_{\alpha \beta \gamma}$ and source flavour. However, it is the K\"ahler moduli that break supersymmetry -
$D_T W \neq 0$ while $D_U W = 0$. As the metric of (\ref{bdmetric}) is block-diagonal, this is equivalent to
$F^T \neq 0$ and $F^U = 0$. The moduli controlling supersymmetry breaking are decoupled from the moduli controlling
flavour, and thus the soft terms generated are naturally flavour-universal. IIB flux models, and in particular the
large-volume models used above, naturally map into the general analysis above,
\bea
\Psi & \Longleftrightarrow & {\hbox{K\"ahler moduli}}, \nonumber \\
\chi & \Longleftrightarrow & {\hbox{Complex structure moduli}}. \nonumber
\eea
We can now understand the calculations of sections \ref{secstc} and \ref{Nondiagonal} as simply
the calculational illustration of this decoupling.

We also note that the above arguments also apply to the
dilaton-dominated 
scenario in heterotic string theory, which is well known to give universality
\cite{bim}. 
In the heterotic string the dilaton has a shift symmetry,
\be
\hbox{Im}(S) \to \hbox{Im}(S) + \epsilon_i,
\ee
which corresponds to a shift in the universal axion and is unbroken in perturbation theory.
This implies the dilaton cannot appear in the tree-level superpotential and thus cannot enter the Yukawa couplings.
If the dilaton dominates the supersymmetry breaking, the supersymmetry breaking sector is again decoupled
from the flavour sector and soft term universality is guaranteed. In
this case the correspondence is
\bea
\Psi & \Longleftrightarrow & {\hbox{Dilaton}}, \nonumber \\
\chi & \Longleftrightarrow & {\hbox{Complex structure moduli}}. \nonumber
\eea

In the IIB context the product matter metric
\be
\tilde{K}_{\alpha \bar{\beta}} = f(T + \bar{T})\tilde{k}_{\alpha \bar{\beta}}(\phi)
\ee
gives soft terms with exact flavour universality.
We noted that we expect this product form to break down at subleading order in $\tau_s$.
As $\tau_s$ controls the gauge coupling, this expansion corresponds to loop corrections.
We would naturally expect these to be flavour sensitive -
loop corrections are sensitive to
particle masses, and thus different flavours should feel these corrections in different ways.
The subleading terms that destroy the product form will also destroy
exact universality, which will only hold at leading order. Another way of stating this is to say that in the presence
only of $\mc{N}=1$
supersymmetry, the decoupling of the $T$ and $U$ moduli will hold only at leading order and will
not be preserved to all orders in perturbation theory.

However these are side issues which do not affect our main point,
that soft term flavour universality is naturally expected in IIB flux models: the complex structure moduli
source flavour, the K\"ahler moduli source supersymmetry breaking, and at leading order the two do not talk to each other.

\subsection{CP violation}
Let us also discuss CP violation in the context of the above soft terms.
The MSSM adds a set of potentially dangerous additional sources of CP violation to
the Standard Model. These can be parametrised as follows \cite{choi92}, \cite{hepph9604387}:
\be
\phi_A  = \{ {\rm arg} \left({A_{\alpha\beta\gamma}\over Y_{\alpha\beta\gamma}}\right)\}, \, \phi_B = \{ {\rm arg}
  B\} ,\, \phi_C = \{ {\rm arg} (M_a)\} .
\ee
The physically relevant combinations $\phi = \{ \phi_A - \phi_C, \phi_B - \phi_C\} $ are
strongly constrained by the measurements of the electric dipole moment, 
$\phi\le 10^{-2} - 10^{-3}.$ 

Let us now consider the soft masses obtained in (\ref{softtermAm}) and
(\ref{ndeq1})-(\ref{ndeq4}). The phase of $A_{\alpha\beta\gamma}/Y_{\alpha\beta\gamma}$ is
aligned with the phase of $F^s/(2\tau_s)$, i.e. the gaugino
mass. Therefore $\phi_A-\phi_C$ vanishes. The same argument can be
applied to deduce that $\phi_B - \phi_C$ also vanishes.
Therefore there are no extra CP violating contributions, and the
experimental constraint on the phases $\phi$ is satisfied in the
scenario under consideration. 

\subsection{Anomaly-Mediated Contributions}
\label{AnomalyMediated}

In this section we compute the magnitude of anomaly-mediated contributions
to gaugino masses following \cite{hepth9911029}
(also see \cite{gaillardnelson}). In that paper a formula
is given for anomaly-mediated contributions to gaugino masses in supergravity,
\be
\label{PoppitzBagger}
m_{1/2} = - \frac{g^2}{16 \pi^2} \left[ (3 T_G - T_R) m_{3/2} - (T_G - T_R) K_i F^i
- \frac{2 T_R}{d_r} (\ln \det K \vert_R '')_i F^i \right].
\ee
Here $K \vert_R ''$ is the K\"ahler metric restricted to the matter fields in representation $R$ (i.e. we only take derivatives
with respect to the matter fields). $d_R$ is the dimension of the representation and $T_R$ is the Dynkin index of the
representation. There is an implicit sum over matter fields and representations in (\ref{PoppitzBagger}). $T_G$ is the Dynkin index
of the adjoint representation, and $3T_G - T_R$ is the coefficient of the $\beta$-function. In minimal anomaly-mediation, the
latter two terms of equation (\ref{PoppitzBagger}) are absent. The $F^i$ are the normal supergravity F-terms
for the moduli fields.

The dependence of the matter metric on moduli fields enters (\ref{PoppitzBagger}) through the third term and in particular
through $ \ln \det K \vert_R ''$. As disussed earlier, we write
\be
K_{C \bar{C}} = \frac{h(\tau_s, \phi) C \bar{C}}{\tau_b^p}.
\ee
We assume there are $d_R$ matter fields. The metric $K \vert_R ''$ is simply the matter metric and is a $d_R \times d_R$ matrix,
with
\be
(K \vert_R '')_{i \bar{j}} = \frac{h_{i\bar{j}}(\tau_s, \phi)}{\tau_b^p},
\ee
where
\be
h_{i \bar{j}}(\tau_s, \phi) = \tau_s^{\lambda} X_{i \bar{j}}(\phi) +
\tau_s^{\lambda - 1} X'_{i \bar{j}}(\phi) + \ldots.
\ee
Then
\be
(\det K \vert_R '') \sim \frac{\tau_s^{\lambda d_R}}{\tau_b^{p \, d_R}} X(\phi),
\ee
where $X(\phi)$ is a complicated function of complex structure moduli. These expressions hold so long as the matter fields
have vanishing vevs.
We then have
\be
\ln \det K \vert_R '' = \lambda d_R \ln \tau_s - p d_R \ln \tau_b + \ldots,
\ee
and
\be
\frac{1}{d_R} (\ln \det K \vert_R '')_i F^i = \frac{\lambda F^s}{2 \tau_s} - \frac{p F^b}{2 \tau_b}.
\ee

We also recall that
\be
K_i F^i = - \frac{3 F^b}{2 \tau_b} \left( 1 + \mc{O}(\mc{V}^{-1}) \right) = 3 m_{3/2} (1 + \mc{O}(\mc{V}^{-1})).
\ee
We can then evaluate the anomaly-mediated contribution to gaugino masses to obtain
\be
m_{1/2} = - \frac{g^2}{16 \pi^2} \left[ 2 T_R (1-p) m_{3/2} - 2 \lambda T_R \left(\frac{F^s}{2 \tau_s} \right) \right].
\ee
This expression has a similar form to those previously encountered in our analysis of soft terms: there is a dominant part,
which cancels for the critical value $p=1$ and a suppressed part depending on the F-term of the small modulus
(we recall this is naturally suppressed by a factor 30 compared to the
gravitino mass). For the actual values $p=1$ and
$\lambda = 1/3$, we obtain
\be
\label{amsbgaugino}
m_{1/2, AMSB} = \frac{g^2}{16 \pi^2} \frac{2}{3} T_R \left( \frac{F^s}{2 \tau_s} \right).
\ee
Because of the no-scale $p=1$ cancellation, this is suppressed compared to the tree-level term by the loop factor
$\frac{g^2}{16 \pi^2}$ and is thus negligible.

Thus, while we may have expected a competition of modulus and anomaly mediation due to the suppressed tree-level gaugino masses,
this does not in fact occur. The anomaly-mediated contributions are doubly suppressed: once by the loop factor, and once
by the same suppression as occurred at tree-level. The root of this double suppression lies in the no-scale structure, which
gives rise to the $p=1$ cancellation.

Similar behaviour will occur for anomaly-mediated contributions to scalar masses. For a no-scale model, these
cancel as for anomaly-mediated contributions to gaugino masses
\cite{randallsundrum,gaillardnelson}. In our models, the no-scale structure is broken
by the F-terms for the small moduli, and these will give rise to anomaly-mediated contributions as in
(\ref{amsbgaugino}). However, as in (\ref{amsbgaugino}) such contributions will be doubly suppressed, once by the
loop factor and once by the small modulus F-term
\footnote{This
  behaviour should also be present in other configurations such as for
  D3 branes in KKLT models that, to leading order, have a no-scale
  structure and cancellations of soft terms will happen for both the
  standard gravity mediation and anomaly mediation.}.

The upshot of these discussions is that, owing to the pseudo-no scale structure,
the anomaly-mediated contributions are suppressed by a loop factor compared to the tree level soft terms
and not compared to the gravitino mass. This implies that they can be consistently neglected in our analysis of
soft terms.

\subsection{Comparison with previous results}

In \cite{hepth0605141} a preliminary calculation of scalar masses was
performed without knowledge of the matter metric. In particular, the
power of the large modulus $\tau_b$ in (\ref{FactorisedK}) was left undetermined.
Denoting this power by $p$, it was there found that
\bea
M_i & = & \frac{F^s}{2 \tau_s} \sim \frac{m_{3/2}}{\ln m_{3/2}}, \\
m_i^2 & = & \left(1 - p \right) m_{3/2}^2 + \frac{F^s \bar{F}^s}{(2
  \tau_s)^2}.
\eea

Since the power $p$ was not known for chiral matter, reference
\cite{hepth0605141} considered the generic value of $p \neq 1$. In
this case scalar masses should be comparable to the gravitino mass
and thus significantly larger than the (logarithmically suppressed)
gaugino masses. This was used for considering universality of
scalar masses since scalar masses could be written as
\be \label{eqn}
m_i^2\ =\ \left[\left(1 - p \right)+
\epsilon_i^2 \right] m_{3/2}^2, \ee
with $\epsilon_i\sim 1/\log\left(M_P/m_{3/2}\right)\sim 1/30$ being
responsible for the breaking of universality. This was understood
by arguing that if the Standard Model lives on a small cycle,
and the stronger source of SUSY breaking is the F-term associated to
the size of the large cycle, then  locality would imply that the
source of non-universality could be associated to the suppressed
value of the F-term corresponding to the small cycle, that is
suppressed.

However the results of \cite{CCQPaper} show that for bifundamental matter actually $p =
1$. While this does not obviate any of the calculations of
\cite{hepth0605141}, it does change the interpretation, as for
$p=1$ the scalar masses are comparable to the gaugino masses
rather than logarithimically larger. For matter with $p \neq
1$, the scalar masses would be logarithmically larger than
the gaugino masses. An example of such matter would be D7 adjont matter.
It would be interesting to find other examples of matter giving $p \neq 1$.

This modifies the discussion of universality. We however have been
able to prove in sections \ref{Nondiagonal} and \ref{Universality} 
that there is still leading-order flavour universality
even for the second term in equation (\ref{eqn}). This is because
the source of non-universality is the flavour
mixing in the Yukawa couplings that have no perturbative dependence on
any of the K\"ahler moduli, large or small.

\section{RG Running and Sample Spectra}
\label{RGrunning}

We now turn to a phenomenological analysis of the large volume models
with soft terms as described in \S \ref{secstc}. Since the soft
terms are computed at the string scale, to obtain the low energy
spectra they need to be evolved to the weak scale using the RG
equations of the MSSM. For this purpose we use the program
\verb+SoftSUSY2.0+ \cite{hepph0104145}. By doing so we are making the
strong theoretical assumption that the
only charged matter content below the string scale is the
MSSM.\footnote{The dilaton-dominated scenario with an intermediate
  string scale was also investigated in \cite{aikq}. In addition to
  the improved software and constraints since \cite{aikq}, this work
  also differs from \cite{aikq} in that the soft terms are derived
  after a process of moduli stabilisation rather than assumed \emph{ad hoc}.} 

The resulting low energy spectra are then subjected to various
experimental constraints. These are the size of the $b\to s\gamma$ branching
ratio, the anomalous magnetic moment of the muon $(g-2)_\mu$ and 
the dark matter relic density $\Omega h^2.$

The average measurement of the $b\to s\gamma$ branching ratio is 
obtained from \cite{heavyflavour}: $BR(b\to s\gamma) = (3.55\pm
0.26) \ti 10^{-4}.$ The theoretical uncertainty in this result is
\cite{hepph0410155} $0.30\ti 10^{-4}$; adding the two errors in
quadrature, we obtain the $1\sigma$ bound, $BR(b\to s\gamma) =
(3.55\pm 0.40)\times 10^{-4}.$

We also impose limits on the new physics contribution to $a_\mu =
(g_\mu-2)/2.$ The average experimental value of $a_\mu$ is
$11659208\times 10^{-10}$ \cite{hepex0401008}. 
Following \cite{pdg}, we use the Standard Model
computation based on data from $e^+ e^-\to {\ \rm hadrons}$, which predicts
$11659186\times 10^{-10}$, rather than that based on tau leptons.
Combining the two results (with respective uncertainties) yields the
$1\sigma$ bound on the non-SM contribution to $a_\mu$, $\delta
a_{\mu} = (22.2\pm 10.2)\times 10^{10}.$ 
It is important to note
that $\delta a_{\mu}$ usually has the same sign as $\mu$, so $\mu>0$
is preferred by experiment.

We also impose bounds on the Higgs mass obtained by the LEP2
collaborations \cite{hepex0306033}. The lower bound is $114.4$ GeV
at the 95\% CL. 
The theoretical computation of the Higgs mass in supersymmetric
scenarios is subject to an estimated error of $3$GeV,
and so we impose $m_h>111$GeV on the result obtained from
\verb+SoftSUSY+.

\begin{figure}
{\fourgraphs{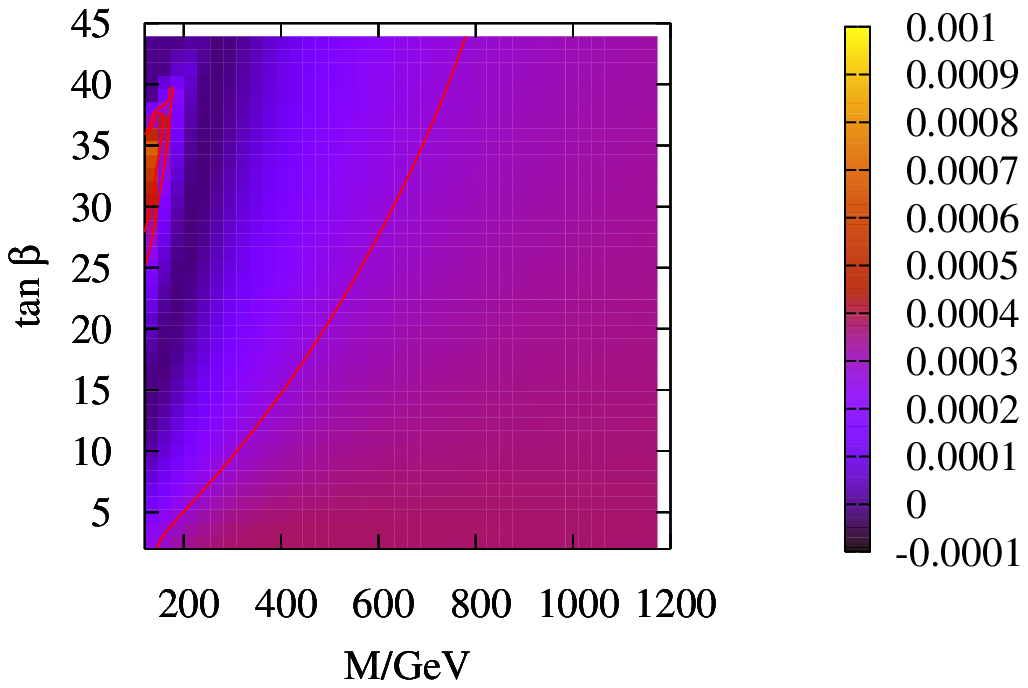}{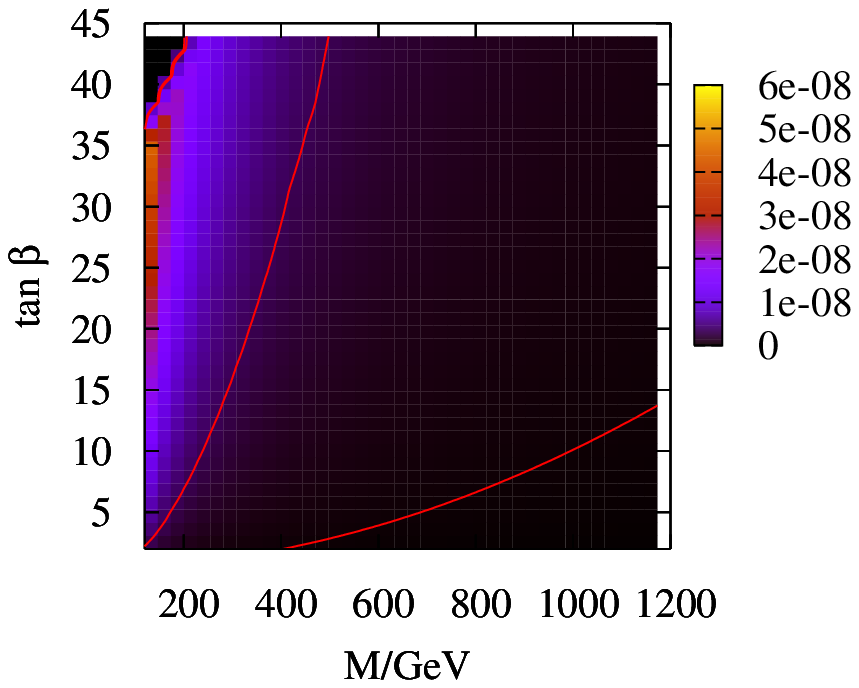}{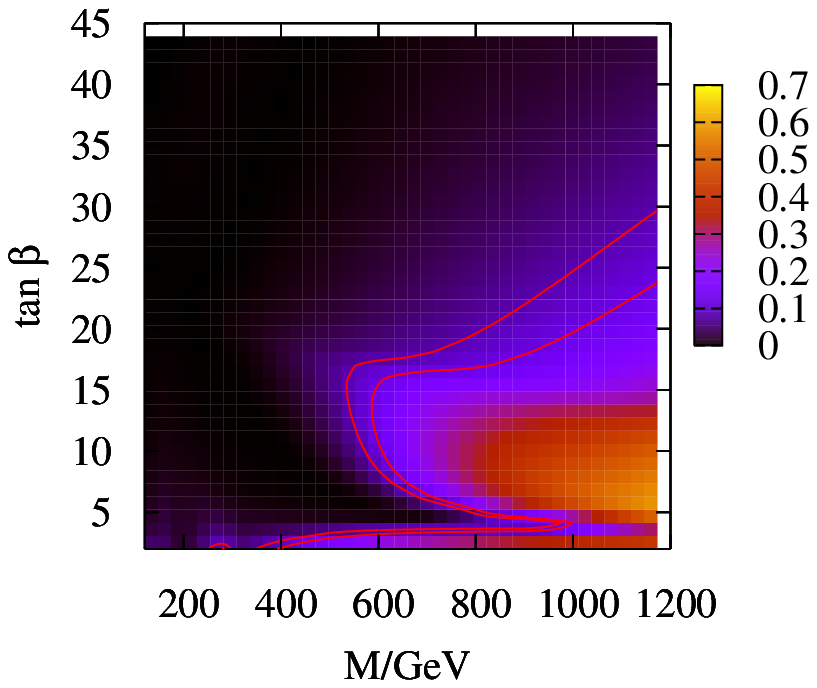}{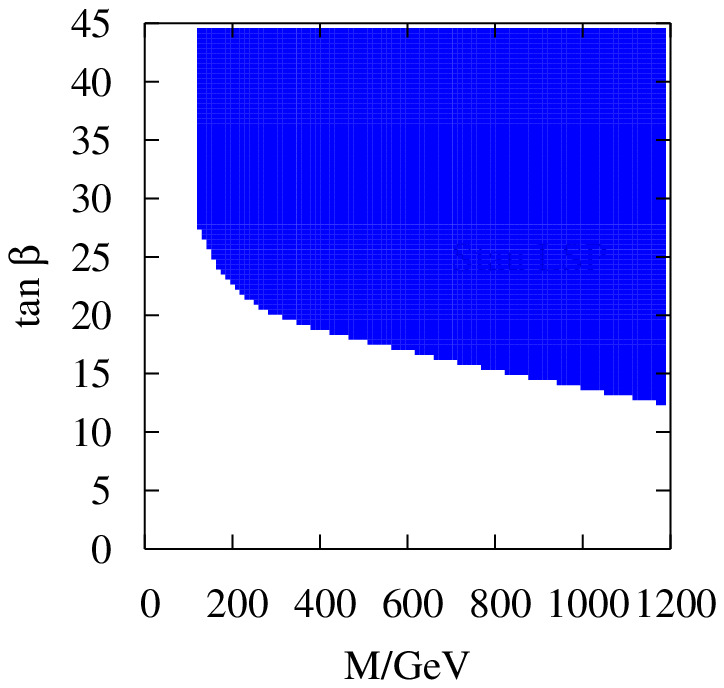}}
\caption{Contour plots of (a) $BR(b\to s\gamma)$ (b) $\delta a_{\mu}$
  and (c) $\Omega h^2$ on $\tan\beta$ and $M$, for $\mu>0$ and $m_s\sim
10^{11}$ GeV. $2\sigma$ bounds are used. The blue region in (d)
denotes that the LSP is stau, the white that it is neutralino.
\label{bsgamma10to11}}
\end{figure}

\begin{figure}
{\fourgraphs{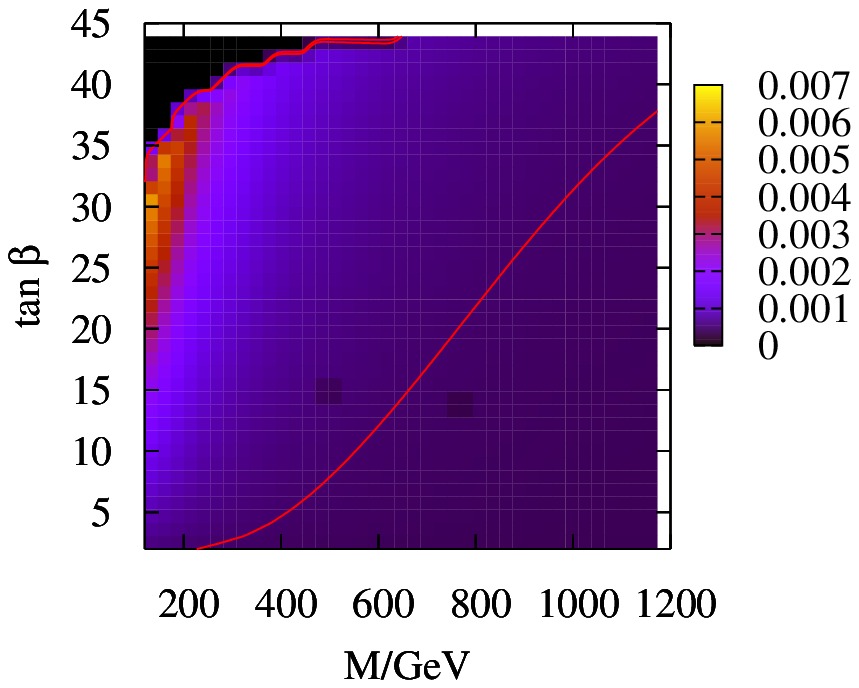}{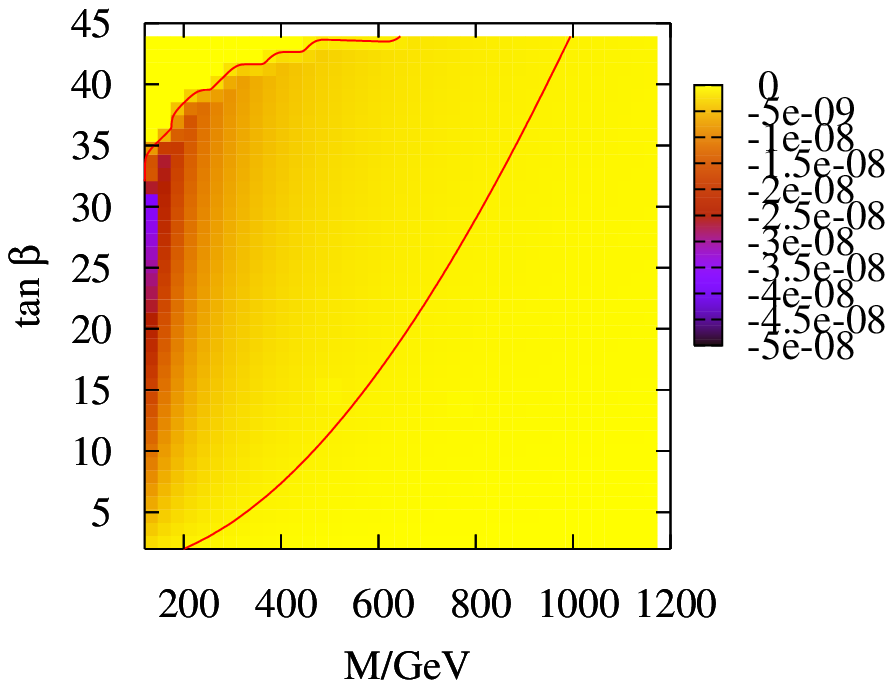}{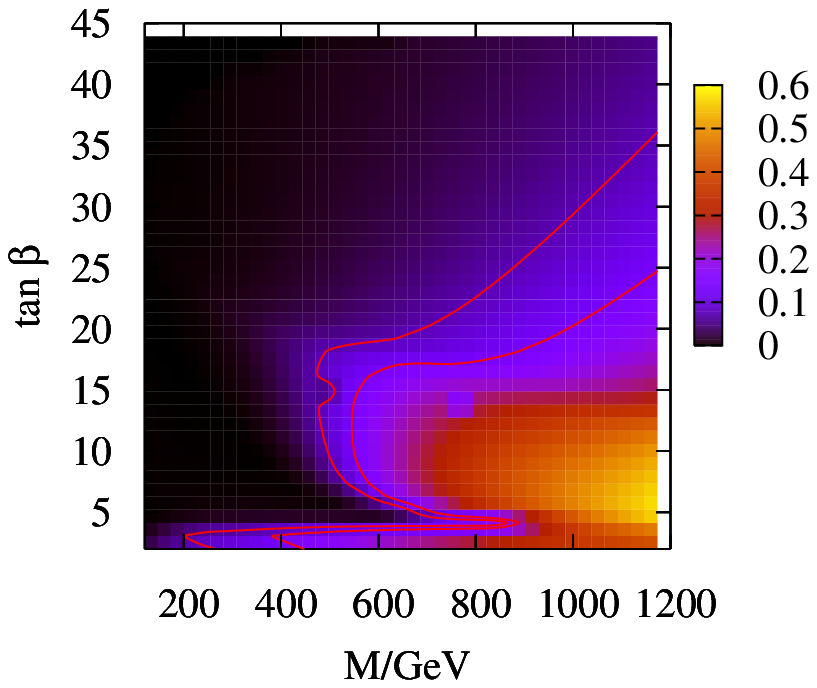}
{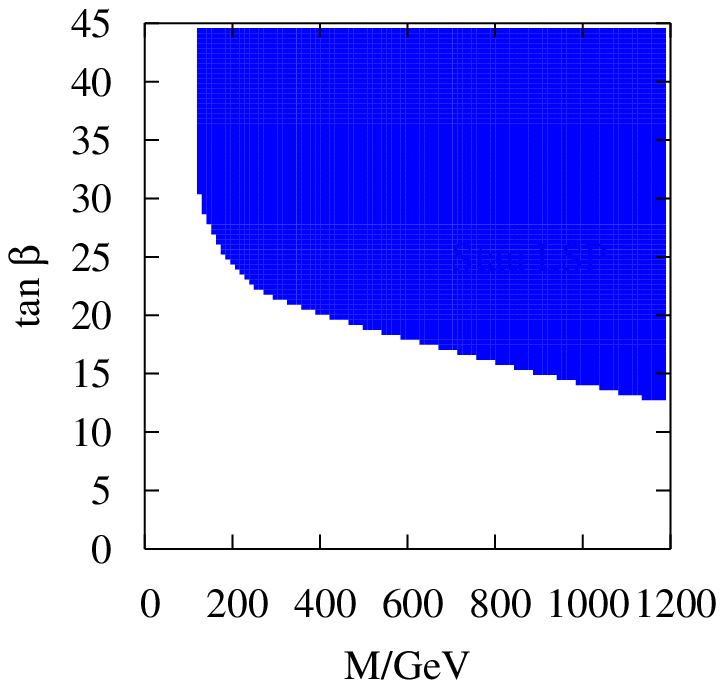}}
\caption{Contour plots of (a) $BR(b\to s\gamma)$ (b) $\delta a_{\mu}$
and (c) $\Omega h^2$ on $\tan\beta$ and $M$ for $\mu<0$ and $m_s\sim
10^{11}$ GeV. Note that $3\sigma$ bounds are used. The blue region in (d)
denotes that the LSP is stau, the white that it is neutralino.
\label{bsgamma10to11muNeg}}
\end{figure}

The WMAP \cite{WMAP} constraint on the relic density of dark matter
particles (at the $2\sigma$ level) is
\begin{equation}
\label{omh2}
0.085 < \Omega h^2 < 0.125.
\end{equation}
We compute the neutralino contribution to dark matter using
\verb+micrOmegas1.3+  \cite{hepph0405253}. The lower bound in
(\ref{omh2}) is only 
applicable if we require that the neutralino is the only contribution
to dark matter. Since there may be other dark matter constituents, such
as axions or for the large volume models the volume modulus,
the upper bound on $\Omega h^2$ is the most relevant.

The values of the Standard Model input parameters used in our
computations are the current central values: $m_t = 171.4$GeV
\cite{hepex0608032}, $m_b (m_b)^{\overline{MS}} =4.25$GeV,
${\alpha}_s (M_Z) = 0.1187$, $\alpha^{-1} (M_Z)^{\overline{MS}} =
127.918$, $M_Z = 91.1187$GeV \cite{pdg}.

\begin{figure}
{\twographsbig{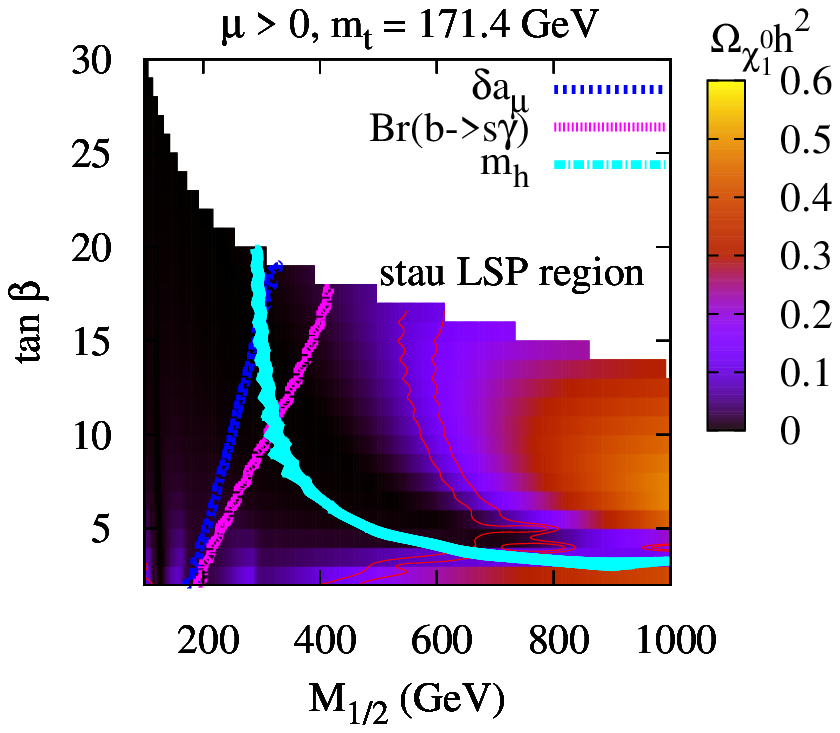}{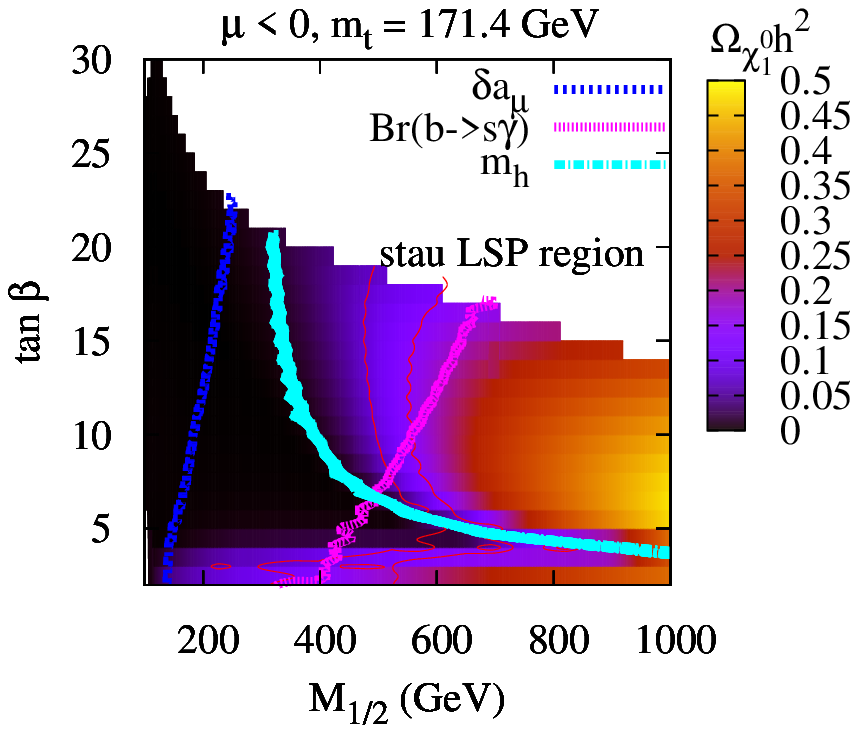}}
\caption{Experimental constraints on the parameter space: If a
  neutralino LSP is to account for the CDM part of the
  Universe, then the allowed regions 
 $ 0.085 < \Omega_{\tilde{\chi}_1^0}h^2 < 0.125 $ for (left) $\mu>0$ and $
  0.075 < \Omega_{\tilde{\chi}_1^0}h^2 < 0.135 $ for (right) $\mu<0$
  are bounded by the red contours. The bounds are
  at $2\sigma$ and $3\sigma$ respectively. The white region
  represents non neutralino (stau) LSP points. The blue and pink
  curves respectively are the $\delta a_\mu$ and $BR(b\to s\gamma)$ 
bounds. The cyan curve is the 111 GeV bound on the Higgs mass.} 
\label{Rgrunfig}
\end{figure}

\begin{figure}
{\twographs{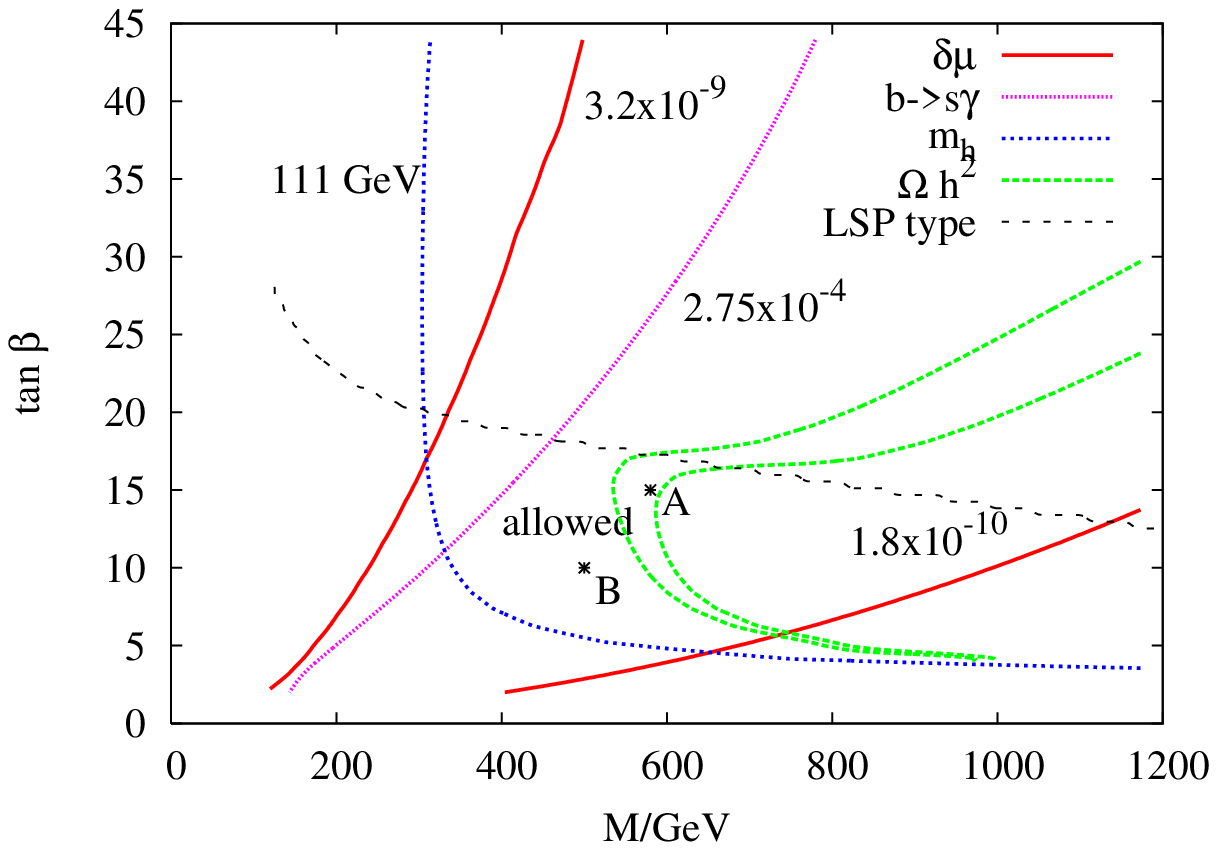}{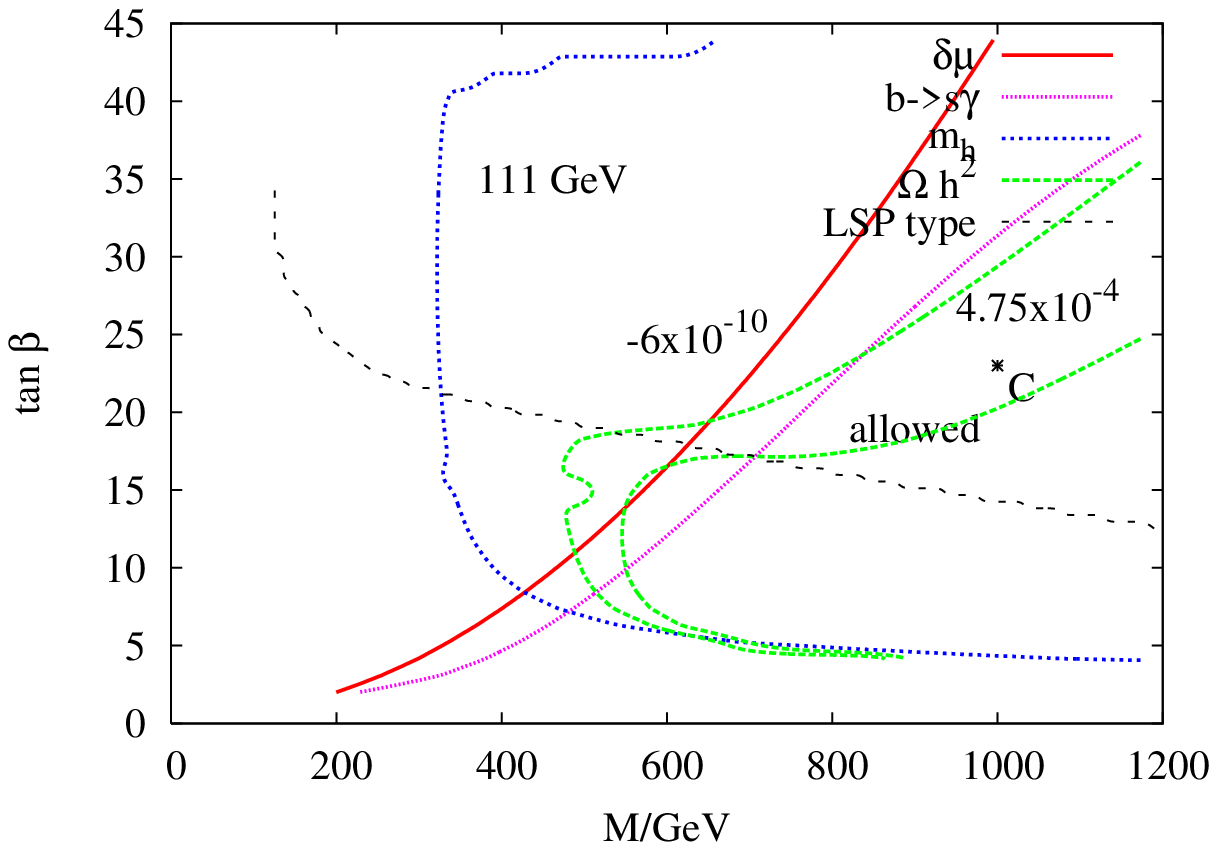}}
\caption{Contour plots of $\delta a_{\mu}$, $BR(b\to s\gamma)$ and
$\Omega h^2$ on $\tan\beta$ and $M$ for (a) $\mu>0$ and (b) $\mu<0$
and $m_s\sim 10^{11}$ GeV. $2\sigma$ bounds are used for $\mu>0$
while $3\sigma$ ones are used for $\mu<0.$ \label{plaincombined}}
\end{figure}

We first neglect the constraint on the  $B$ soft term and scan over various
values for the gaugino mass $M$ and $\tan\beta.$ The
\verb+micrOmegas1.3+ package, interfaced with
\verb+SoftSUSY2.0+ via the SUSY Les Houches Accord
\cite{hepph0311123}, is used to compute the $b\to s\gamma$ branching
ratio, $\delta a_{\mu}$ and $\Omega h^2.$ 
The constraints are presented on separate plots in figures
\ref{bsgamma10to11} and \ref{bsgamma10to11muNeg}. 

Figures \ref{bsgamma10to11}, \ref{bsgamma10to11muNeg} also show regions in 
the $M-\tan\beta$ plane with stau LSP. The existence of a stable charged LSP is
experimentally ruled out \cite{heavyisotope} and therefore if such a
scenario is to be considered, it must include R parity violation to
allow the LSPs to decay before nucleosynthesis.

We also scanned the $M-\tan\beta$ parameter space for potentially dangerous
charge and colour breaking (CCB) minima (see e.g. \cite{softreview,
  hepph9707475}). This was done by investigating the unbounded
from below UFB3 direction
in field space. As described in \cite{hepph9507294}, UFB3 is the most
restrictive of the UFB and CCB constraints. Other less
restrictive CCB minima were also investigated. In the region of
interest we found no such minima, for both the $\mu>0$ and $\mu<0$
cases.

The combined results involving all the experimental constraints and displaying
stau LSP regions are shown in figures \ref{Rgrunfig}, \ref{plaincombined}.

We can see that in the $\mu>0$ regime there is a very restricted throat-like
region in the $M-\tan\beta$ plane which satisfies all the experimental
constraints (including dark matter) and has a neutralino LSP. 
If the lower bound on $\Omega h^2$ is not imposed the region is much larger.
In the $\mu<0$ regime the experimentally allowed region is entirely 
stau LSP, so one is forced to consider R-parity violating models
(for a recent discussion see \cite{ben}.

\begin{figure}[ht]
\begin{center}
   \begin{minipage}[t]{6.0cm}
    \includegraphics[angle=90, width=1.3\textwidth]{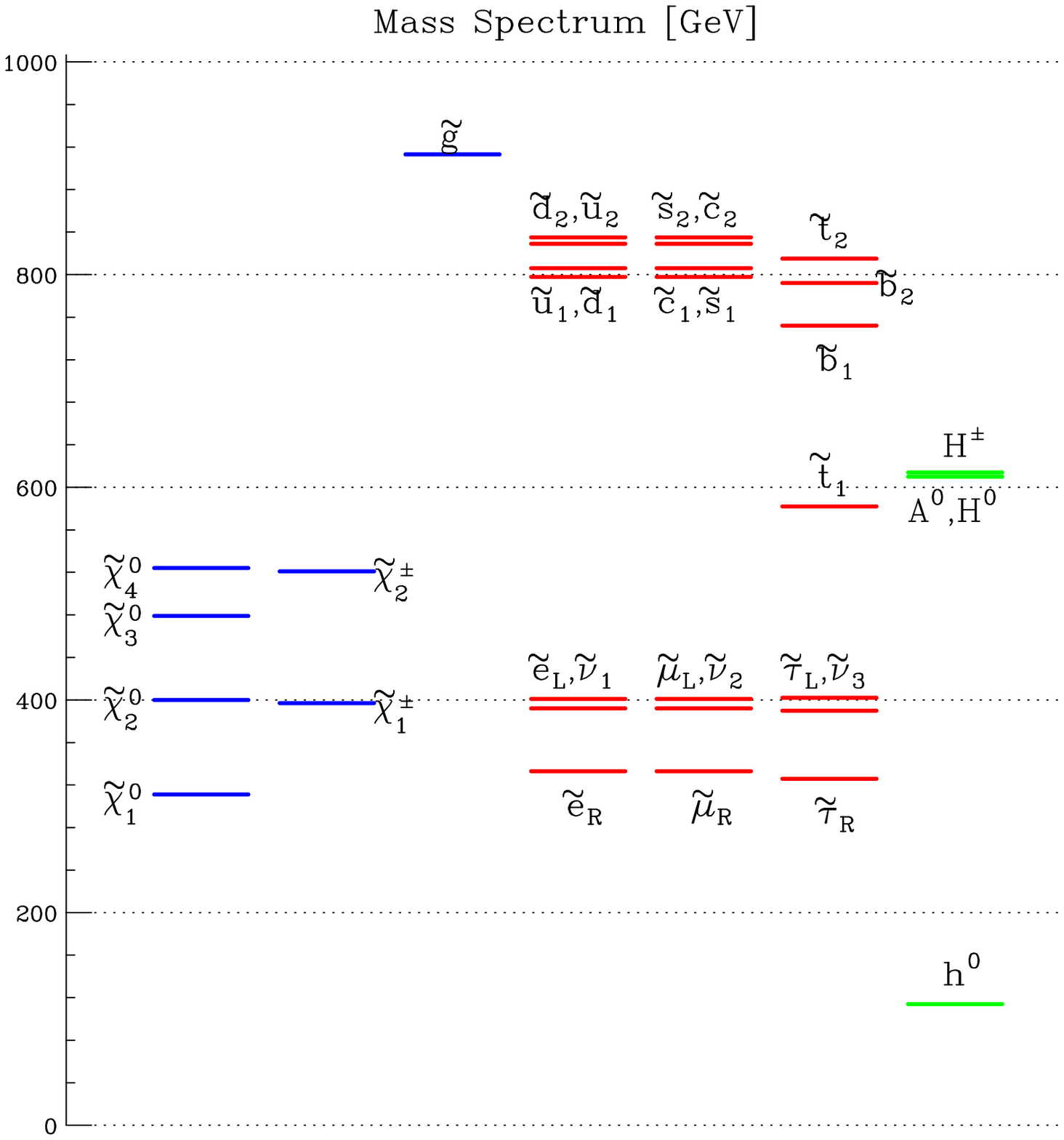}
   \end{minipage}
\end{center}
  \caption{\footnotesize{Pictorial representation of sample mass spectra in
  allowed $\mu > 0$ region. The numerical values are shown in column B
  of the sample spectra table.}}
\label{spectraB}
\end{figure}

\begin{figure}[ht]
\begin{center}
   \begin{minipage}[t]{6.0cm}
    \includegraphics[angle=90, width=1.30\textwidth]{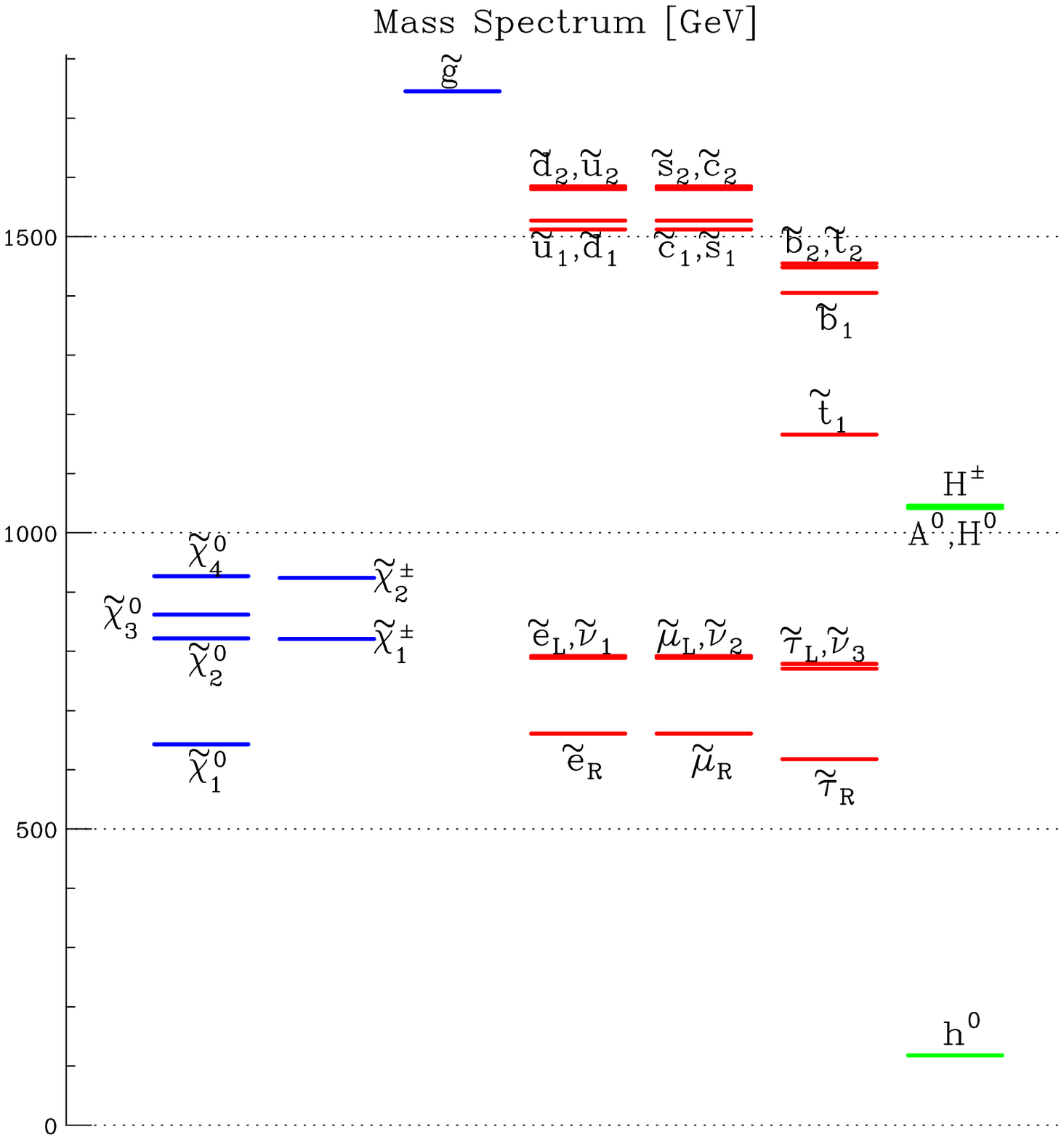}
   \end{minipage}
\end{center}
  \caption{\footnotesize{Pictorial representation for sample mass spectra in
  $\mu < 0$ and stau LSP region. The numerical values are shown in
  column C of the sample spectra table.}}
\label{spectraC}
\end{figure}

Table \ref{Spectra} shows a few sample points, one with $\mu>0$
from the allowed region in figures \ref{Rgrunfig} and
\ref{plaincombined} and one with 
$\mu<0$ with stau LSP, which pass all the experimental constraints
we imposed. The corresponding spectra computed using \verb+SoftSUSY+
are also shown. The spectra B and C are also pictorially represented in 
figures \ref{spectraB} and \ref{spectraC}.

\begin{figure}
{\twographs{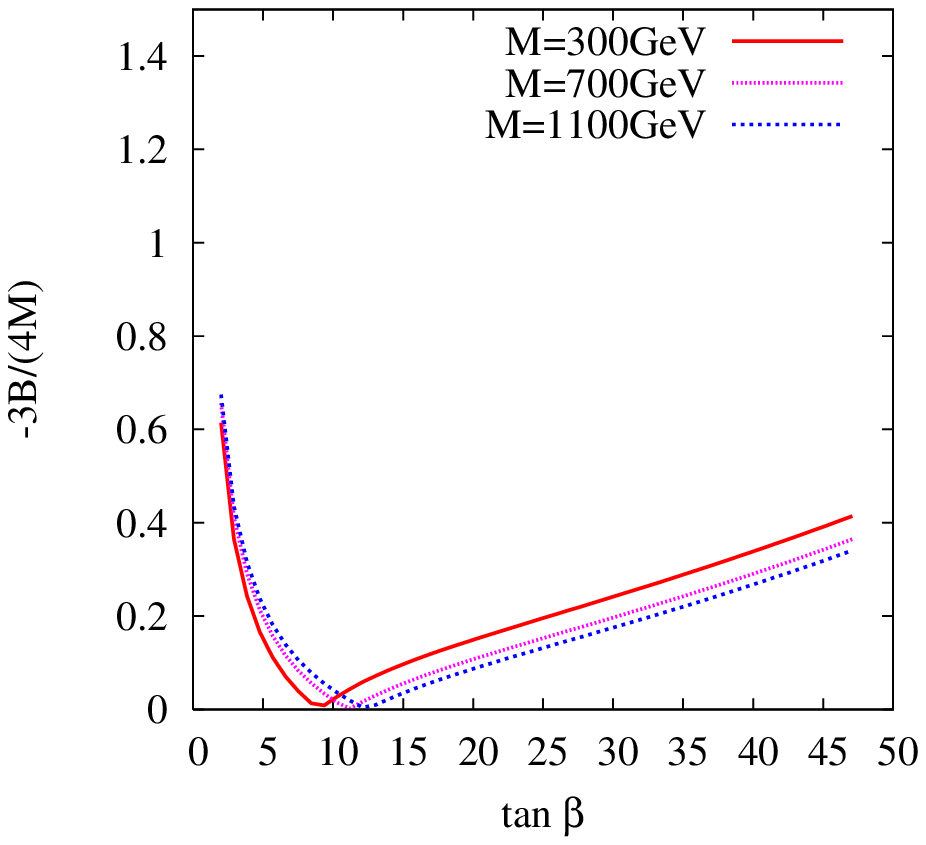}{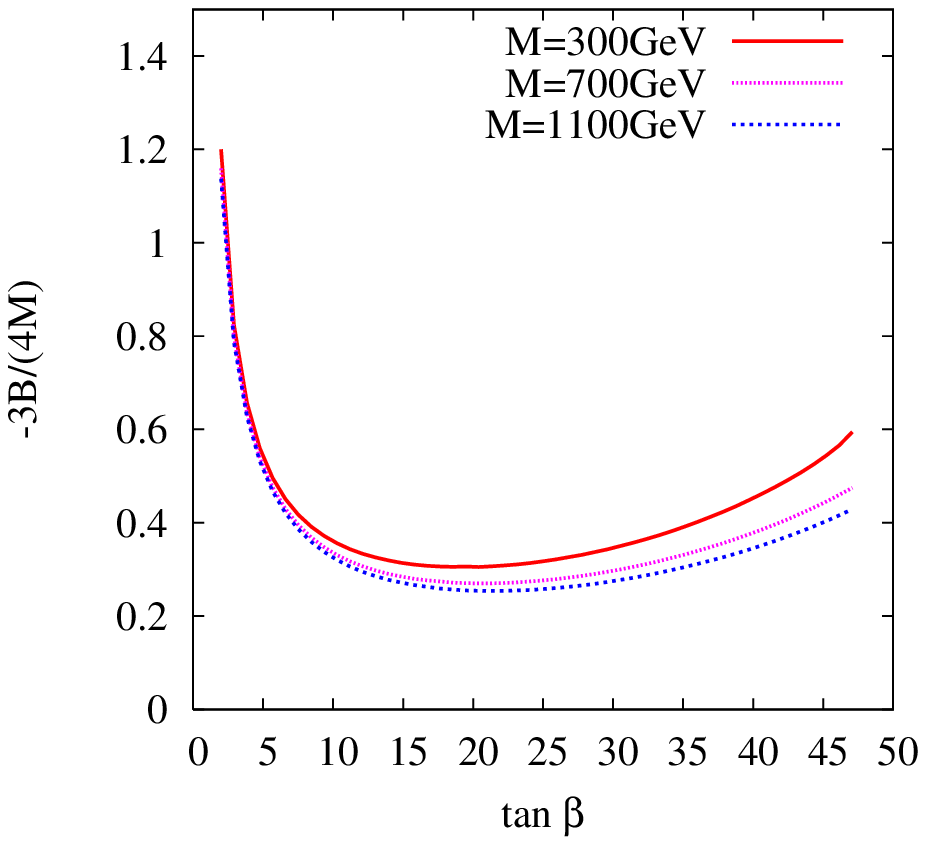}} \caption{The ratio
$B/(-4M/3)$ over a range of values of $\tan\beta$ for (a) $\mu>0$
and (b) $\mu<0$ and $m_s\sim 10^{11}$ GeV. \label{btomratio}}
\end{figure}

We next attempt to also impose the high scale constraint $B=-4M/3.$
This turns out to be impossible to satisfy for $\mu>0$, as Figure
\ref{btomratio} indicates. For $\mu<0$ it can only be satisfied in
the very low $\tan\beta$ region. However, for those values of
$\tan\beta$ the Higgs mass is below the LEP bound.

Therefore we are forced to assume that there is some mechanism at work
responsible for generating the correct $B$-term, e.g. a low energy
coupling of the form $\alpha N H_1 H_2$ with a gauge invariant
scalar $N$ which obtains an appropriate vev.

One necessary caveat on all the above results is that they are all obtained in
the dilute flux approximation, where we have neglected the
contribution of magnetic fluxes to the gauge couplings. A further
caveat is that there may be additional vector-like matter at the TeV
scale that will alter the running of the soft terms (for explicit
string models where this effect has been studied see \cite{alt}). The results
are therefore necessarily approximate and there is a certain amount of
`theoretical error' in the soft terms used.\footnote{In order
to estimate the changes that our assumptions may have, we also
computed the spectrum for a model with the spectrum of the MSSM plus
extra doublets such that gauge unification is achieved at the
intermediate scale. The spectrum takes a similar structure as the
one shown here, with the scalar masses increasing by $10-20$ GeV,
and the lightest neutralino becoming roughly $40$ GeV lighter; this
has the interesting consequence of expanding the neutralino LSP region
in figure \ref{Rgrunfig}.} 
 In particular, we have
used universal gaugino masses even though the Standard Model gauge
couplings are not unified at the intermediate scale.
In the large
volume model, we interpret this lack of unification as due to the
magnetic fluxes. These fluxes will also modify the matter metrics, and
in general contribute corrections to all the soft terms considered
here. The magnitude of these corrections can be roughly estimated by
the fractional non-universality of the $SU(2)$ and $SU(3)$ gauge
couplings at the intermediate scale.
The above results should all be understood as leading order results in
the dilute flux approximation. 
It would be interesting to
perform a careful study of the extent to which the inclusion of magnetic fluxes would
alter the results in this section.

\begin{table}
\begin{center}
\begin{tabular}{|c|c|c|c|c|}
\hline
& A & B & C\\
\hline
$m_s$ & $10^{11}$ & $10^{11}$ & $10^{11}$ \\
\hline
$\tan\beta$ & 15 & 10 & 23\\
\hline
$M$ & 580 & 500 &  1000\\
\hline
${\rm sgn} \mu$ & + & + & - \\
\hline \hline
$\tilde{e}_L, \tilde{\mu}_L$ & 464  & 401 &  792\\
\hline
$\tilde{e}_R, \tilde{\mu}_R$ & 386  & 333 &  661\\
\hline
$\tilde{\tau}_L$ & 463  & 402 & 779\\
\hline
$\tilde{\tau}_R$ & 369  & 326 & 618\\
\hline
$\tilde{u}_1, \tilde{c}_1$ & 924  & 806 & 1527\\
\hline
$\tilde{u}_2, \tilde{c}_2$ & 951  & 829 & 1580\\
\hline
$\tilde{t}_1$ & 679 & 582 &  1166\\
\hline
$\tilde{t}_2$ & 958 & 815 &  1448\\
\hline
$\tilde{d}_1, \tilde{s}_1$ &  915  & 798 &  1512\\
\hline
$\tilde{d}_2, \tilde{s}_2$ &  958  & 835 &  1585\\
\hline
$\tilde{b}_1$ &  859  & 752  &  1405\\
\hline
$\tilde{b}_2$ &  903  & 792 &  1455\\
\hline
$\chi_1^0$ &  364  & 311 &  643\\
\hline
$\chi_2^0$ &  469  & 400 &  822\\
\hline
$\chi_3^0$ &  541  & 479 &  862\\
\hline
$\chi_4^0$ &  587  & 524 &  927\\
\hline
$\chi_1^{\pm}$ &  467  & 397 & 821\\
\hline
$\chi_2^{\pm}$ &  584  & 521 & 924 \\
\hline
$A_0, H_0$ &  679  & 610 & 1042\\
\hline
$H^{\pm}$  &  684  & 614 &  1046\\
\hline
$\tilde{g}$ &  1048  & 913 &  1745\\
\hline
$\tilde{\nu}_{1,2}$ & 456 & 392 &  789\\
\hline
$\tilde{\nu}_3$ & 451  & 390 &  771\\
\hline
$h$ &  116  & 114 & 118\\
\hline \hline
$B(b\to s\gamma)/10^{-4}$ & $3.3$ &  $3.4$ & $4.42$ \\
\hline
$\delta a_{\mu}/10^{-10}$ & $7.9$ &  $7.0$ & $-4.3$ \\
\hline
$\Omega h^2$ & 0.12 & 0.01 & --- \\
\hline
\end{tabular}
\end{center}
\caption{ Sparticle spectra for two sample points with $\mu>0$ and
$\mu<0$. All masses are in GeV. \label{Spectra}}
\end{table}

\section{The Simplest  KKLT Scenario}
\label{kkltl}

Let us now briefly recall the one-modulus KKLT scenario. In this case
there is only one $T$ field and the volume is given by $\mc{V}=(T+ T^*)^{3/2}$. Here the flux
superpotential is required to be very small, $W_0\sim 10^{-13}$, in
order to address the hierarchy problem. The minimum of the
scalar potential is supersymmetric AdS and supersymmetry is only
broken by the lifting term, which unfortunately is the least
understood ingredient of the KKLT scenario. This has an important
effect on the structure of soft breaking terms, leading to the
overall suppression by the factor $\ln(M_P/m_{3/2})$. Soft terms have been
computed for general values of the modular weight in the matter
K\"ahler potential. However the phenomenological analysis has mainly been
performed for vanishing modular weight which is standard for non-chiral 
D7 adjoint matter.

The expression for soft masses found in  \cite{choi} takes the form:
\begin{eqnarray}
\label{KKLTmasses}
m_\lambda^2 = (1-n_\lambda)\left| {F^T\over{(T+T^*)}}\right|^2 - {1\over{32\pi^2}} 
{d\gamma_\lambda\over
d \log \mu} \left|{F^\Phi\over \Phi_0}\right|^2 +\nonumber \\ 
\left( {1\over8} \sum_{\mu\nu} (3-n_\lambda - n_\mu - n_\nu) |Y_{\lambda \mu\nu}|^2 - 
{1\over2} \sum_a l_a T_a (C^\lambda) g_a^2 \right)\cdot \nonumber \\
\left( {F^T\over{(T+T^*)} } \left( {F^{*\Phi}\over {4\pi^2 \Phi^*_0}}  \right)
+ {F^{*T} \over {(T+T^*)}} \left( {{F^{\Phi}\over{4\pi^2 \Phi_0}}} \right)
 \right).
\end{eqnarray}
Here $\Phi$ is the conformal compensator field, $b_a$ are the one-loop beta 
function coefficients, $g_a$ the gauge couplings, $\gamma_\lambda$ are the
anomalous dimensions of the matter fields $C^\lambda$, and $T_a$ are group theory factors. 
Finally, $n_\lambda$ is the power of $T+T^*$ in the expansion of the K\" ahler potential 
in matter fields,
\be
K = -3 \log (T+T^*) +{C^\lambda C^{*\lambda}\over{(T+T^*)^{n_\lambda}}}. 
\ee
The F-terms can be written as 
$$
F^\Phi \sim 16\pi^2 M_s, \qquad F^T/(T+T^*)=\alpha M_s.
$$
The parameter $\alpha$ depends on the shape of the potential used to lift the supersymmetric
AdS vacuum to a de Sitter one, but is typically in the range $4-6$ \cite{choi}.
The soft terms (\ref{KKLTmasses}) can be rewritten as
\be
m_\lambda^2 = M_s^2 ((1-n_\lambda) \alpha^2 - \dot{\gamma}_\lambda + 
2\alpha (T+T^*) \partial_T \gamma_\lambda),
\ee 
with $\dot{\gamma}_\lambda = 8\pi^2 \partial\gamma_\lambda/\partial \log \mu.$

In the paper \cite{falkowski} the value $n_\lambda=0$ corresponding to a field in the adjoint
representation was used. However, in \cite{CCQPaper} it was 
shown that the correct modular weight for a chiral field is actually 
$n_\lambda=2/3$. 
Using the explicit expressions for $\dot{\gamma}$ and $(T+T^*) \partial_T \gamma_i$ 
(see \cite{falkowski}), one then finds boundary conditions at the GUT scale for the 
first two generations of particles:
\begin{eqnarray}
m^2_L &\sim& (-1-2\alpha + {\alpha^2/3}) M_s^2,\nonumber\\
m^2_E &\sim& (-2-\alpha + {\alpha^2/3}) M_s^2,\nonumber\\
m^2_Q &\sim& (2-4\alpha + {\alpha^2/3}) M_s^2,\nonumber\\
m^2_U &\sim& (1-3\alpha + {\alpha^2/3}) M_s^2,\nonumber\\
m^2_D &\sim& (2-3\alpha + {\alpha^2/3}) M_s^2.
\end{eqnarray}
Requring all the masses to be positive (in order to prevent tachyons and/or charge
and colour breaking minima from appearing) gives a lower bound on $\alpha$, 
$\alpha\ge 11.5.$ Such values are difficult to obtain in the KKLT
scenario.


%


A full reanalysis of the phenomenology of this scenario is beyond
the scope of this article. Note that combining gravity and anomaly
mediation also leads to some ambiguity of the soft terms. Even though
gaugino masses have been fully determined in all generality
\cite{hepth9911029}, scalar masses  have a Planck scale physics
dependence that is not usually under control. Fortunately for the
large volume case the cancellation was due to the standard
cancellation of soft terms for no-scale supergravity that has been
established  in \cite{randallsundrum, gaillardnelson} and this
ambiguity is under control.\footnote{The ultraviolet effects can be
controlled for instance by regularising with Pauli-Vilars fields that appear in
the no-scale combination (we thank E. Poppitz, M.K. Gaillard and B.
Nelson for helpful discussions on this point).} Furthermore, since
in the KKLT scenario the value of $T$ is not very large, $\alpha'$
corrections should play an important role. It would be interesting
to consider these corrections using the results of \cite{vijayper}.

\section{Conclusions}
\label{Conclusions}

In this paper
we have been able to go a long way towards developing a concrete
model of low-energy supersymmetry breaking for chiral matter fields
from D7 branes. Notice that D3 branes were considered in \cite{hepth0505076,
aqs} but there was a need to fine tune in order to obtain low-energy
supersymmetry breaking. Previous discussions of soft terms for
D7 branes used the matter metrics for non-chiral adjoint
fields since the K\"ahler metric for chiral fields was unknown.
Using the recent results of \cite{CCQPaper}, we were able to find the
soft supersymmetry breaking terms for chiral fields also, which are
the ones most relevant for phenomenology.

We have found several interesting results.
\begin{enumerate}

\item
There is a small hierarchy between the gravitino mass and all the
soft breaking terms. This is given by the suppression factor
$\epsilon^{-1}=\log(M_P/m_{3/2})$. This allows the gravitino to be
more than one order of magnitude heavier than the scalars and
gauginos. Without having to fine-tune $W_0$, the large-volume scenario solves the
hierarchy problem with an intermediate string scale $m_s \sim 10^{11}
\hbox{GeV}$. The scenario also naturally generates an axion decay
constant within the allowed window \cite{hepth0602233}.
We are then
led to a hierarchy of scales $m_s \sim 10^{11} {\rm
GeV} \gg m_{3/2}\sim 10 {\rm TeV}> m_{soft}\sim 500 {\rm GeV}$.

\item
For  general non-diagonal matter metrics we were able to extract a
remarkably simple structure of soft terms, depending only on an
overall scale $M_i\sim \epsilon m_{3/2}$ and a parameter $\lambda$
which is the modular weight of the corresponding matter fields with
respect to the small cycles,
 \bea \label{lamb}
M_i & = & \frac{F^s}{2 \tau_s}, \nonumber \\
m_{\alpha} & = & \sqrt{\lambda} M_i, \nonumber \\
A_{\alpha \beta \gamma} & = & - 3\lambda  M_i, \nonumber \\
 B & = & - \left(\lambda +1\right)  M_i. \eea
 This produces a well
motivated scenario of soft SUSY breaking at low-energies which can be
subjected to a detailed phenomenological analysis.

\item
The structure of the soft terms provides an elegant explanation of
universality. The origin of universality can be traced to the fact
that the sector that controls flavour (complex structure moduli) is
decoupled, at leading order, from the sector that breaks
supersymmetry (K\"ahler moduli). It is promising to have a solution
to probably the most serious problem of gravity mediated
supersymmetry breaking. Note that universality is only approximate
in our scheme, as it is valid only to first order in an expansion of
the K\"ahler moduli. This is interesting because it predicts the
breaking of universality at next order in a gauge coupling
expansion.

\item
For the simplest case, corresponding to the Standard Model gauge and
matter sectors coming from magnetised D7 branes wrapping the same
cycle, $\lambda=1/3$. The structure of breaking terms is essentially
the same as the dilaton domination scenario and only differs in the
value of the soft terms compared to the gravitino mass and the
expression for the $B$ term. This is quite curious since the
origin of these two mechanisms is completely different. Dilaton
domination was considered very special in the heterotic case due to
its simplicity, universality and potential connection with finite
theories \cite{ibanez, jackjones}. Furthermore, its simplicity
allowed it to be subject to phenomenological constraints, such as the
absence of charge and colour breaking, that ruled this scenario out
for a string
scale of $M_{GUT}$. Fortunately these constraints were satisfied if the
string scale is intermediate \cite{aikq} as indeed happens with our
case. 
Note that the difference with the previous studies is
that in the past the value of the string scale was put in by
hand but now it is derived after a well defined process of moduli
stabilisation. Furthermore in the past there were no explicit models
in which dilaton domination was derived (see also the recent
discussion in \cite{gomezreinoscrucca}).

\item
Even though the soft breaking terms are suppressed with respect to
the gravitino mass, the anomaly mediated contribution is further
suppressed. The main reason for suppression in both cases is the
underlying no-scale structure of our scenario. In pure no-scale models
anomaly mediated soft terms vanish. Since no-scale is
only approximate in the large volume models, the AMSB soft terms are non-vanishing but they are
suppressed compared to the minimal AMSB soft breaking terms. Recall that for
the no-scale model anomaly mediation also gives vanishing soft terms
and therefore its contribution is doubly suppressed (by the no-scale
approximate cancellation and the standard loop suppresion).
Therefore, contrary to the one-modulus KKLT case, anomaly mediation
does not play a role in our scenario.

\item
The scenario provides interesting low-energy implications. 
Regarding CP violation, since the $A$ terms are proportional to the
gaugino masses, dangerous CP violating phases are not present (see
for instance \cite{choi92}). Second, there are regions in parameter space
satisfying the standard constraints of charge and
colour breaking minima, neutral LSP,  $BR(b\rightarrow s\gamma)$, $(g-2)_\mu$
and dark matter. 

\item
The simplest one-modulus KKLT scenario also gets modified due to the
knowledge of the chiral field matter metric. Even though this may not change
the broad properties of the model, the detailed  phenomenology is
changed due to the particular value of the modular weight, being
$2/3$ and not $0$ as has been assumed in the
phenomenological discussions in \cite{choi,falkowski, mirage}.

\end{enumerate}

There are several open questions that deserve further study. It
would be interesting to further explore all the phenomenological
implications of this scenario, analysing collider signatures in a similar fashion to the pioneering
work of \cite{gordy}. We have also neglected the contribution of fluxes to
the gauge couplings and therefore also to the gaugino masses. This corresponds
to magnetic fluxes being diluted in the corresponding
4-cycle. Our analysis in this paper has all been done in the dilute
flux approximation. In addition to gaugino masses, the fluxes will
also enter the K\"ahler metrics. For this reason it would be very
interesting to extend this work by going beyond the
limit of large cycle volumes and dilute fluxes. In general it should
be remarked  that the running to low-energies can only be seen as an
indication of the implications of our soft terms, since some of its
details depend on the precise structure of the realistic string
model, including spectrum and couplings which are model dependent.

We have considered explicitly a minimal model in which the K\"ahler
metric depends only on the volume and the corresponding `small'
modulus. A more general case would include dependence on other small
moduli.

One issue with the above large-volume scenario is cosmological.
As has been emphasised in \cite{hepth0505076, hepth0605141}, our scenario is such
that all but one of the K\"ahler moduli get masses of order
$m_{3/2}/\epsilon$. These moduli decay very rapidly and cause no
cosmological problem.
However the modulus
corresponding to the size of the large 4-cycle is lighter than the
gravitino and extremely long lived. 
This may cause cosmological problems which need to be addressed
\cite{hepph9308292, hepph9308325}.

Finally, this scenario has a dichotomy about preferred values of the
string scale. Either we have the GUT scale motivated by gauge
unification and inflation, or we have the intermediate scale,
motivated by the solution of the hierarchy problem in 
gravity-mediated scenarios, as well as being the natural scale for axions to solve
the strong CP problem. We have taken for our analysis the second
option, motivated by the fact that to have the GUT scale we need
very small values of $W_0$. We consider this to be a very strong fine
tuning, which diminishes the significance of low-energy supersymmetry in
addressing the hierarchy problem. Natural values of $W_0\sim 1$ select
the intermediate scale. If the use of very small
values of $W_0$ can be justified, an extension of our analysis to include GUT scale
string theory may be considered. This would be a replacement for the
GUT scale fluxed MSSM models considered in
\cite{hepph0408064},\cite{hepph0502151},\cite{aqs}, 
where the adjoint field metrics were used. However, it is well known that the
entire $M-\tan\beta$ parameter space in the GUT scale dilaton domination
scenario suffers from the presence of CCB minima \cite{hepph9701357},
rendering the 
scenario rather unattractive.

It would also be interesting to search for scenarios in which the good properties of both the GUT
and intermediate scale could be at work.

\section*{Acknowledgements}

We acknowledge useful conversations with 
 S.P. de Alwis, E. Bassouls, M. Berg, C.P.
Burgess, M.K. Gaillard, 
D. Grellscheid, L.E.  Ib\'a\~nez, S. Kachru, G. Kane, S.
Kom, P. Kumar, B. Nelson, H.P. Nilles, E. Poppitz. We especially
 thank B. Allanach and  D. Cremades for many
interesting discussions. SA is supported by a Gates Fellowship. JC
is supported by EPSRC and Trinity College, Cambridge. FQ is
partially supported by PPARC and a Royal Society Wolfson award. KS
is supported by Trinity College, Cambridge. SA thanks the organisers of
YETI'06-BSM, IPPP Durham. FQ thanks KITP,  Santa Barbara and the
organisers of the `String Phenomenology' program, where part of this
work was developed and which was funded in part by the
 National Science Foundation under Grant No. PHY99-07949.

\end{document}